\documentclass{aa}
\usepackage{multirow}

\usepackage[T1]{fontenc}
\usepackage{txfonts}
\usepackage{amsmath}
\usepackage{amsfonts}
\usepackage{xcolor}
\usepackage{url}
\usepackage{bm}
\usepackage{graphicx}
\usepackage{amssymb}
\usepackage{soul}
\usepackage{booktabs}
\usepackage{array}
\usepackage[utf8]{inputenc}
\inputencoding{latin1}
\inputencoding{utf8}
\usepackage{appendix}

\def\lsim{ \lower .75ex\hbox{$\sim$} \llap{\raise .27ex \hbox{$<$}} }
\def\gsim{ \lower .75ex \hbox{$\sim$} \llap{\raise .27ex \hbox{$>$}} }

\newcommand{\grad}{\nabla}

\newcommand{\bi}{\begin{itemize}}
\newcommand{\ei}{\end{itemize}}

\usepackage[colorlinks=true,linkcolor=blue,citecolor=blue]{hyperref}

\begin{document}

\title{Do multifrequency polarimetric observations of BL Lac \\ rule out a hadronic origin for its X-ray emission?}
\titlerunning{Proton-synchrotron radiation and polarization in BL Lac}

\author{
F. Tavecchio\inst{1}
\and
F. Bolis\inst{2,1}
\and
E. Sobacchi\inst{3,1}
\and
S. Boula\inst{1}
\and
A. Sciaccaluga\inst{1}
}
\authorrunning{Tavecchio et al.}

\institute{
INAF -- Osservatorio Astronomico di Brera, Via E. Bianchi 46, I-23807 Merate, Italy\\
\email{fabrizio.tavecchio@inaf.it}
\and
DiSAT, Universit\`a dell’Insubria, Via Valleggio 11, I-22100 Como, Italy
\and
Gran Sasso Science Institute, Viale F.~Crispi 7, I-67100 L’Aquila, Italy
}
\date{}

\voffset-0.4in



\abstract{Recent multifrequency polarimetric observations of the eponymous blazar BL Lac reveal an extremely large degree of polarization in the optical band (average of $25\%$, reaching $45\%$), together with a small degree of polarization in the X-ray band ($\lesssim 7\%$). This has been interpreted as evidence that the X-rays are produced through inverse Compton emission by relativistic electrons, thus ruling out alternative models based on hadronic processes. Here we revisit the observational evidence, interpreting it in a framework where
the observed radiation is entirely produced through synchrotron emission. Electrons produce the radio-to-optical component, and protons produce the X-rays and the gamma rays.
We determined the jet magnetic fields from a magnetohydrodynamic model of magnetically dominated stationary axisymmetric outflows, and show that the X-ray emission from the protons is naturally less polarized than the optical emission from the electrons. The model parameters required to reproduce the multifrequency polarimetric observations are fully compatible with blazar jets.
}


\maketitle
\boldsymbol{}

\section{Introduction}

Relativistic jets produced by active galactic nuclei (AGNs) represent ideal laboratories for addressing several topics related to black hole physics, particle acceleration, and high-energy (and possible multi-messenger) emission \citep[e.g.,][]{Blandford19}. Jets are best studied in blazars, where a favorable geometrical alignment leads to the relativistic amplification of the nonthermal radiation produced in the jet by relativistic particles \citep{romero17}. Blazars are characterized by a double-humped spectral energy distribution (SED), with the first hump peaking in the IR-soft-X-ray band and the high-energy hump peaking at gamma-ray energies (e.g., \citealt{fossati98}).
There is general consensus that the first hump is produced by leptons (electrons or pairs) through synchrotron emission.  According to the standard framework \citep[e.g.,][]{sikora94,ghisellini98}, the same particles up-scatter soft radiation, forming the high-energy hump through the inverse Compton (IC) mechanism. A competing scenario assumes instead that the high-energy hump is produced by relativistic hadrons, either through direct proton-synchrotron emission or through synchrotron emission by the by-products of hadronic reactions (see \citealt{SolZech22} for a review). The hadronic scenario is supported by the potential association of some blazars with high-energy neutrinos \citep[e.g.,][]{Boettcher19}.

The Imaging X-ray Polarimetry Explorer (IXPE) satellite \citep{weisskopf22} enables multifrequency polarimetric observations of blazars that, in principle, can offer unique clues as to the physical processes acting in these sources. In particular, it has been anticipated that the polarimetric properties of the X-ray emission of blazars can be used to determine the emission mechanism responsible for the high-energy hump \citep[e.g.,][]{Zhang17,peirson22}. This can be done in low- and intermediate-synchrotron-peaked (LSP and ISP) blazars (\citealt{Abdo10}), where the X-rays trace the rising portion of the high-energy hump. The general idea is that in single-zone models with nearly uniform fields, hadronic processes lead to a high polarization of the radiation in this band because X-rays are produced through synchrotron emission, either by high-energy protons or by the by-products of hadronic reactions \citep{ZhangBoettcher13}. On the other hand, leptonic models lead to a lower polarization because the X-rays are produced through IC emission. The IC radiation would be unpolarized when the seed photons are thermal and isotropic, such as photons produced in the central regions of the AGN \citep{Bonometto73}, and would have a low degree of polarization ($\lesssim 50\%$ of the degree of polarization of the seed photons) when it is produced through the synchrotron self-Compton (SSC) mechanism \citep{Poutanen94,Krawczynski12}. The comparison between the polarization in the optical band (produced through synchrotron radiation by high-energy electrons) and in the X-ray band would therefore be a powerful diagnostic of the emission mechanism at work.

\cite{Agudo25} report multifrequency polarimetric observations of BL Lac, a well-studied LSP/ISP-type blazar. The optical emission, which belongs to the soft tail of the low-energy SED component, showed an exceptionally high (and variable) degree of polarization (average of $25\%$, reaching 45\%). IXPE obtained a quite constraining upper limit of $\sim 7\%$ for the degree of polarization in the X-ray band, which tracks the rising part of the high-energy SED peak. \cite{Agudo25} conclude that these measurements strongly support the leptonic scenario, in which the X-rays are produced through IC scattering by the same electrons that produce the radio-to-optical component through synchrotron emission. This result strengthens previous similar results by IXPE for BL Lac \citep{middei23b,Peirson23} and other sources of the LSP class \citep{Marshall24}. A follow-up modeling of the source was carried out by \cite{liodakis25}, who conclude that, regardless of the jet composition and the assumed emission model, IC scattering from relativistic electrons dominates the X-ray regime. Their study considered four scenarios for the high-energy spectral component: a pure SSC model, a SSC + external Compton leptonic model, a hybrid model involving SSC and hadronic processes, and a purely hadronic model. 

Given the relevance of the topic, it is important to more thoroughly investigate the hadronic scenarios, which are disfavored by the studies of
\cite{Agudo25} and \cite{liodakis25}. Here we discuss a scenario in which the high-energy component is produced by high-energy hadrons (protons in the following) through synchrotron emission \citep[e.g.,][]{Aharonian2000,Zech17, liodakis20}.
We determined the jet magnetic fields from a magnetohydrodynamic model of magnetically dominated stationary axisymmetric outflows \citep[e.g.,][]{Vlahakis2004,Lyubarsky2009}.
The degree of polarization of synchrotron radiation depends on the shape of the jet, the size of the emission region, and the slope of the energy distribution of the emitting particles \citep{Bolis+2024}. We emphasize that the dependence on the slope of the particle distribution is stronger than in the textbook case of a uniform field. The low degree of polarization in the X-ray band is naturally produced by the joint effect of (i) the hard slope of the underlying proton energy distribution and (ii) the larger region encompassed by the slow-cooling protons with respect to the fast-cooling electrons responsible for the optical emission.

The paper is organized as follows. In Sect.~\ref{sec:model} we present the model and the calculation of the polarization, and in Sect.~\ref{sec:discussion} we discuss the results.
Throughout the paper, the following cosmological parameters are assumed: $H_0=70{\rm\;km\;s}^{-1}{\rm\; Mpc}^{-1}$, $\Omega_{\rm M}=0.3$, and $\Omega_{\Lambda}=0.7$.

\section{The model}
\label{sec:model}

\subsection{Model set-up}

Our aim is to show that multifrequency polarization observations of BL Lac can be explained by a model where the X-rays are produced through the proton-synchrotron mechanism.
Our model follows the work of \cite{Bolis+2024}, which discusses the polarization of synchrotron radiation from magnetically dominated jets.
The main result of that work is the following. The degree of polarization has a strong dependence on the slope of the energy distribution of the emitting particles (see also \citealt{Bolis+2024b}). This feature can be important to explain the strong chromaticity of the observed degree of polarization in high-synchrotron peaked blazars.
Here, we extend the model of \cite{Bolis+2024}, including the emission from a population of high-energy protons that fill a region of non-negligible thickness.

\begin{figure}[t]
    \hspace{-0.4 truecm}
    \vspace{-0.8 truecm}
      \includegraphics[width=1.1\linewidth]{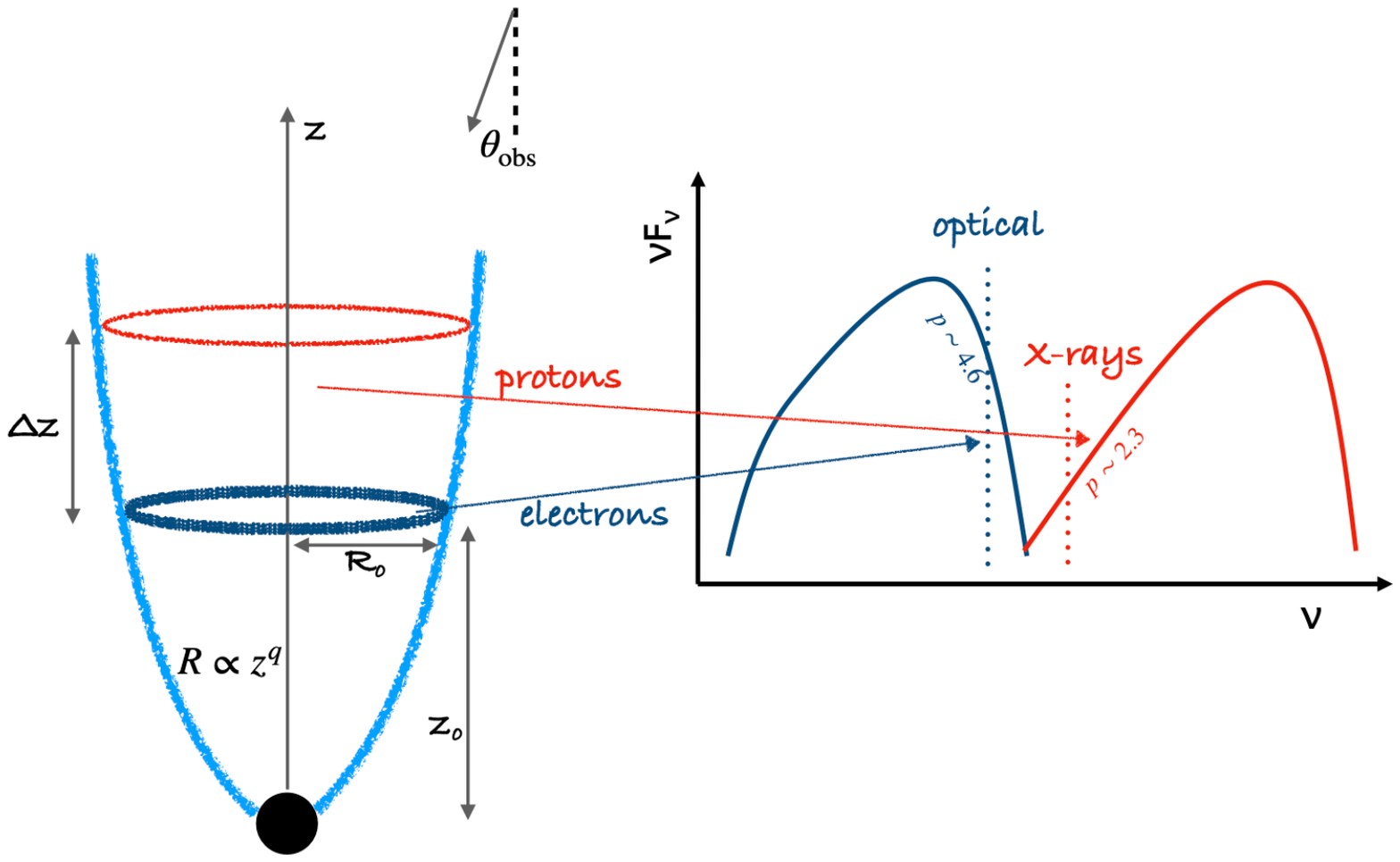}
    \caption{Sketch of the scenario discussed in the text. The jet has a parabolic shape and carries a helical magnetic field. Particles are accelerated at some distance ($z_0$) from the black hole. Electrons, which have a very short cooling time, produce the optical emission in a thin shell close to the acceleration zone. Protons, instead, are in the slow-cooling regime and are advected by the flow up to a distance $\Delta z\sim z_0$ (determined by the adiabatic energy losses). Protons produce the high-energy component of the SED, including the X-rays.}
    \label{fig:cartoon}
\end{figure}

Our scenario is sketched in Fig.~\ref{fig:cartoon}. The transverse radius of the jet, $R_0$, and the distance from the central black hole, $z_0$, are related by $R_0\propto z_0^q$. The jet carries a helical magnetic field (see below for more details).
At some distance $z_0\sim 10^{17} {\rm\; cm}$ from the black hole, particles (electrons and protons) are accelerated to relativistic energies by an unspecified acceleration mechanism and subsequently cool, emitting synchrotron radiation. 
We assumed that the first component of the SED is produced by the electrons and the second component by synchrotron radiation from protons. To produce photons at the peak of the high-energy hump, which is located around $E\sim 100{\rm\; MeV}$, protons have energies on the order of \citep{Aharonian2000}\footnote{Standard notation denotes quantities in the observer frame, whereas primed quantities refer to the proper frame of the fluid.}
\begin{equation}
    \gamma^{\prime}_{\rm p, max}\simeq \left( \frac{4\pi m_p c E}{3ehB^{\prime}\delta} \right)^{1/2}\simeq 3\times 10^8 B^{\prime -1/2}_1 \delta_1^{-1/2}\;,
    \label{eq:gammap}
\end{equation}
where $e$ is the electron charge, $h$ is the Planck constant, $c$ the speed of light, and $m_{\rm p}$ the protons mass. We use the standard notation $B^{\prime}=10\;B^{\prime}_1{\rm\; G}$ and $\delta=10\;\delta_1$, where $B^{\prime}$ is the magnetic field and $\delta$ is the Doppler factor. In the following, we assume fiducial values $B^{\prime}=10{\rm\; G}$ and $\delta=10$. We also assume a viewing angle $\theta_{\rm obs}=0.1{\rm\; rad}$, which implies a bulk Lorentz factor $\Gamma=10$.

The magnetic field is sufficient to confine the protons up to the energy required to produce the high-energy peak, which is $E^{\prime}_{\rm p,max}=\gamma^{\prime}_{\rm p, max} m_{\rm p} c^2= 3\times 10^{17}{\rm\; eV}$ (assuming a jet transverse radius $R\sim10^{16}{\rm\; cm}$, the classical Hillas limit \citep{Hillas} gives $E^{\prime}_{\rm p}\lesssim eB^{\prime}R \sim 10^{19}{\rm\; eV}$). Moreover, one can easily check that the strong magnetic field suppresses the SSC emission, which provides a negligible contribution to the observed emission (see Appendix \ref{sec:SSC}). Finally, photomeson and Bethe-Heitler losses of protons are also negligible with respect to synchrotron\footnote{Using Eq.~(5) of \cite{Murase12} we found that the $p\gamma$ cooling time is about $10^3$ times longer than the synchrotron cooling time for protons emitting at $100 {\rm\; MeV}$.}.

For the adopted strength of the magnetic field, the protons that produce the X-ray emission (at $5 {\rm\; keV}$, the center of the IXPE band) have a Lorentz factor $\gamma^{\prime}_{\rm p}\simeq 4\times 10^6$ (given by Eq.~\ref{eq:gammap} with $E=5{\rm\; keV}$), while the electrons responsible for the optical emission have a Lorentz factor $\gamma^{\prime}_{\rm e}\simeq 10^3$.
The cooling length of the emitting electrons and protons is an important quantity in our model. Electrons are characterized by very efficient radiative losses and, therefore, by a very short cooling length ($l_{\rm e,cool}$):
\begin{equation}
    l_{\rm e,cool} = c \frac{\gamma^{\prime}_{\rm e}}{\dot{\gamma^{\prime}_{\rm e}}}\Gamma = \frac{6\pi m_{\rm e} c^2}{\sigma_{\rm T} {B^{\prime}}^2 \gamma^{\prime}_{\rm e}} \Gamma\simeq 3\times 10^{15} \; {\rm cm}\ll z_0\;,
\end{equation}
where $\sigma_{\rm T}$ is the Thomson cross section and $m_{\rm e}$ the electron mass. On the other hand, protons are in the slow-cooling regime and, therefore, their energy losses (even at the highest energies) are dominated by adiabatic expansion. The typical length scale of adiabatic cooling, $l_{\rm p,cool}=\Delta z\simeq 3z_0/2q$, is on the order of a few $z_0$ (see Appendix \ref{sec:adiabatic}). Therefore, while electrons emit optical radiation in a very thin slice, coincident with the acceleration region, the X-ray emission is produced by protons, advected by the plasma, that fill a much larger volume (see Fig.~\ref{fig:cartoon}). As we show below, this is important to further reduce the polarization of the X-ray emission.

For the jet parameters assumed above, the energy flux associated with the protons is on the order of $P_{\rm p}=3\times 10^{45}{\rm\; erg\; s}^{-1}$ (see Appendix \ref{sec:lp}). The magnetic field carries an energy flux about ten times larger, $P_{\rm B}=3\times 10^{46}{\rm\; erg\; s}^{-1}$. The jet is thus dominated by the Poynting flux, consistent with our assumptions. The total jet power is close but below the Eddington luminosity of the central supermassive black hole, which is $L_{\rm Edd}=6.5\times 10^{46}{\rm\; erg\; s}^{-1}$ for a black hole mass $M_{\rm BH}\simeq 5\times 10^8 M_{\odot}$ \citep{Capetti10}.

\subsection{Polarization}
\label{sec:polariz}

In the following, we sketch the polarization degree calculation, which is presented in detail by \cite{Bolis+2024, Bolis+2024b}. We determined the magnetic fields of the jet from a magnetohydrodynamic model of magnetically dominated
stationary axisymmetric outflows confined by an external medium \citep{Lyubarsky2009}. We assumed that the pressure profile of the external medium decreases as a power law, $\mathcal{P}_{\rm ext}  \propto z_0^{-\kappa}$, where $z_0$ is the distance from the black hole. In this case, the transverse radius of the jet, $R_{0}$, scales as $R_0 \propto z_0^{q}$. The parameter $q$ depends on the external pressure profile. The jet has a parabolic shape for a wind-like medium ($\kappa=2$), with $1/2<q<1$.

Adopting cylindrical coordinates, the electromagnetic fields in the observer frame can be written as
\begin{align}
\label{eq:Efield}
\mathbf{E} & = E_{R} \hat{\mathbf{R}} + E_{z} \hat{\mathbf{z}}\\
\label{eq:Bfield}
\mathbf{B} & = B_{R} \hat{\mathbf{R}} + B_{\phi} \hat{\bm{\phi}} +B_{z} \hat{\mathbf{z}} \;,
\end{align}
where\begin{align}
\label{eq:Ecomp}
E_{R} & = \Omega R B_{\rm p} \cos \Theta\;, & E_{z} & = -\Omega R B_{\rm p} \sin \Theta \\
\label{eq:Bcomp}
B_{R} & = B_{\rm p}\sin \Theta\;, & B_{z} & = B_{\rm p} \cos \Theta \;.
\end{align}
In this section the speed of light was set to $c=1$. The angular velocity, $\Omega$, and the poloidal magnetic field, $B_{\rm p}$, are independent of $R$. The transverse radius of the jet is given by $\Omega R_{0} = 3^{1/4} (\Omega z_0)^{q}$.
The jet opening angle, $\Theta$, and the toroidal magnetic field are given by \citep{Lyubarsky2009, Bolis+2024}
\begin{align}
 \label{eq:angle}
 \Theta & =  q \; \frac{R}{R_{0}} \frac{3^{1 / 4q}}{\left(\Omega R_{0} \right)^{\left(1-q\right) / q}}\\
\label{eq:BminusE}
 \frac{B^{2}_{\phi} - E^{2}}{B^{2}_{\rm p}} & = 3^{-1 + 1 / 2q} \; q \left(1-q \right)  \frac{\left(\Omega R\right)^{4}}{\left(\Omega R_{0}\right)^{2 / q}} \;.
\end{align}
We assumed that the bulk velocity of the fluid, $\mathbf{v}$, coincides with the drift velocity, as appropriate for Poynting-dominated outflows: $\mathbf{v}=\mathbf{E}\times\mathbf{B}/B^2$. The corresponding bulk Lorentz factor is $\Gamma=(1-v^2)^{-1/2}=(1-E^2/B^2)^{-1/2}$. 
The value of $\Omega R_{0}$ was determined from the condition $\Gamma(R_0)=10$, which corresponds to the typical Lorentz factor of AGN jets.

We assumed that the distribution of the emitting particles (electrons and protons) in the fluid proper frame is isotropic in momentum. We also assumed that the distribution is a power law in energy, $N_{\rm e,p}^{\prime}(\gamma_{\rm e,p}^{\prime})= K_{\rm e,p} \gamma_{\rm e,p}^{\prime -p}$, where $K_{\rm e,p}(R,z)$ is the proper particle number density and $\gamma_{\rm e,p}^{\prime}$ is the particle Lorentz factor.

The linear degree of polarization, $\Pi,$ and the electric vector position angle, $\Psi,$ of the synchrotron radiation from the entire emission region (assumed to be unresolved), are respectively given by\footnote{Since we are dealing with ultra-relativistic particles, we neglected circular polarization and set $V=0$.}
\begin{align}
\label{PI}
\Pi &= \frac{\sqrt{Q^{2} + U^{2}}}{I} \\
\label{Psi}
\tan 2 \Psi &= \frac{U}{Q} \;, \quad \rm{where} \; \; 0 < \Psi < \pi \;.
\end{align}
The Stokes parameters $I$, $Q$, and $U$ are calculated in Appendix \ref{sec:Stokes}. We had to define the particle number density of electrons and protons, $K_{\rm e,p}$. As in \cite{Bolis+2024}, we considered two scenarios: $K_{\rm e,p}=K_0$, which corresponds to a constant number density in the proper frame of the fluid, and $K_{\rm e,p}\propto B^{\prime 2} = B^2/\Gamma^2$, which corresponds to constant magnetization.

\begin{figure*}
    \centering
    \includegraphics[width=0.4\linewidth]{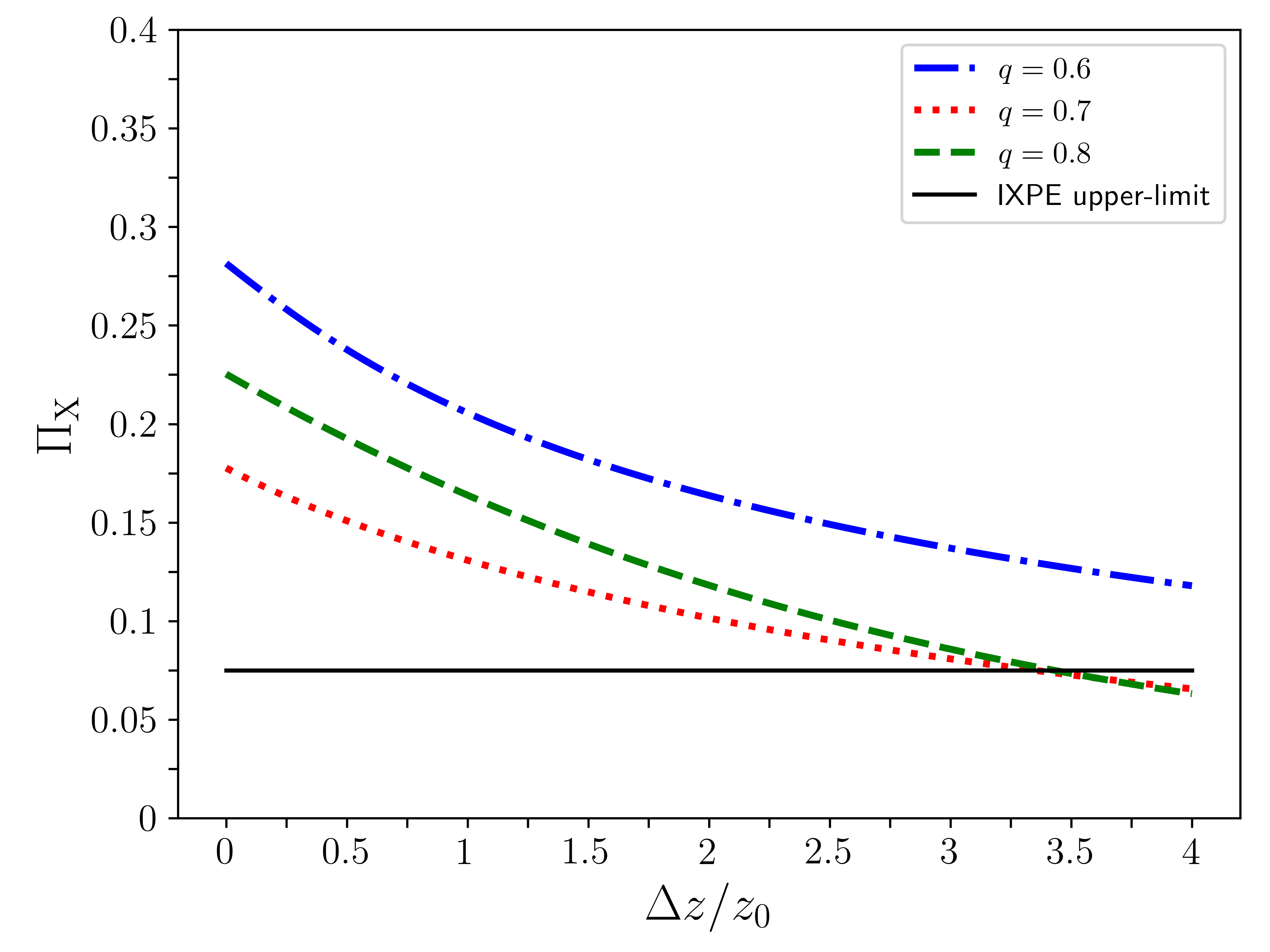}
      \includegraphics[width=0.4\linewidth]{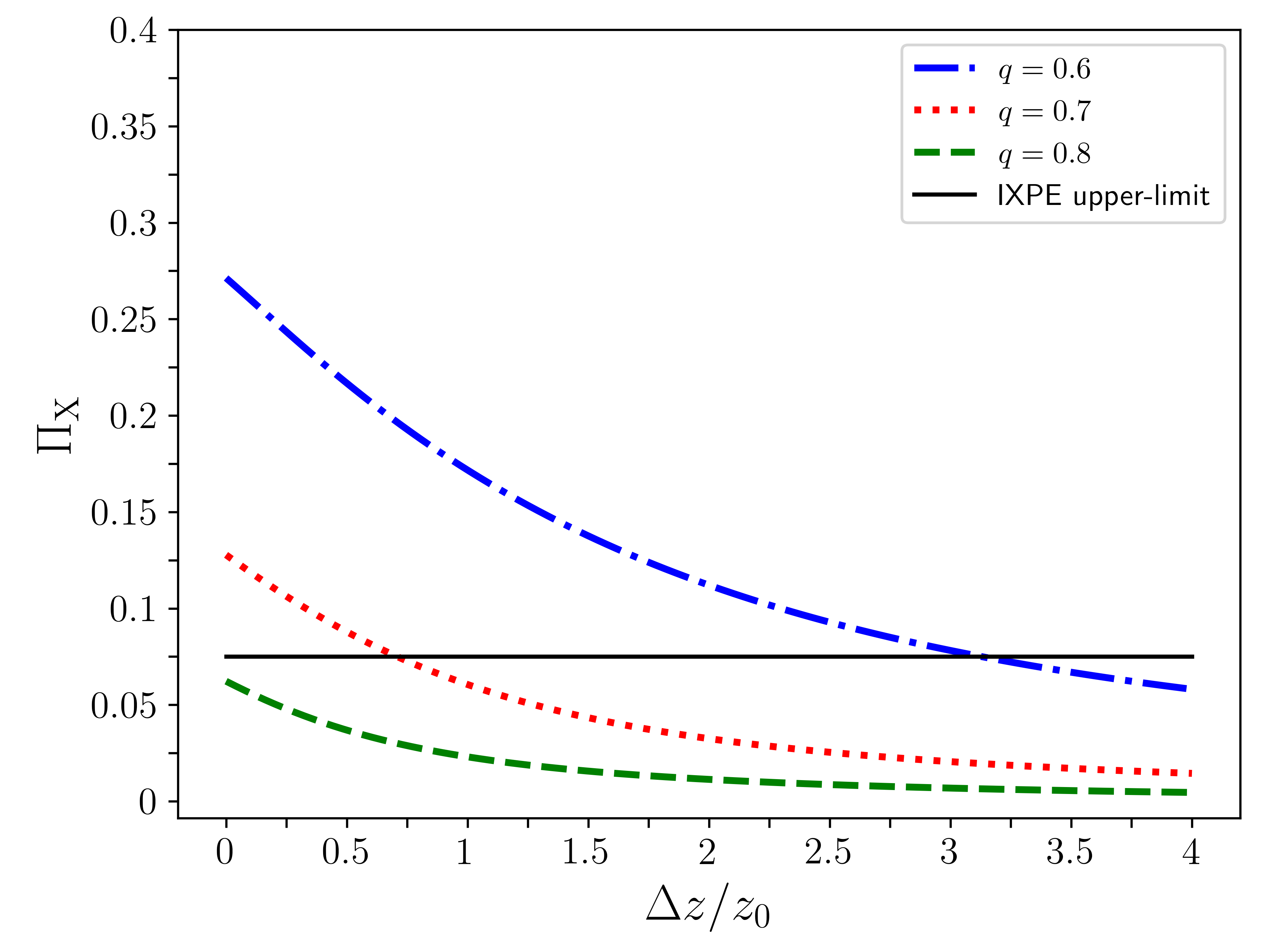}
    \includegraphics[width=0.4\linewidth]{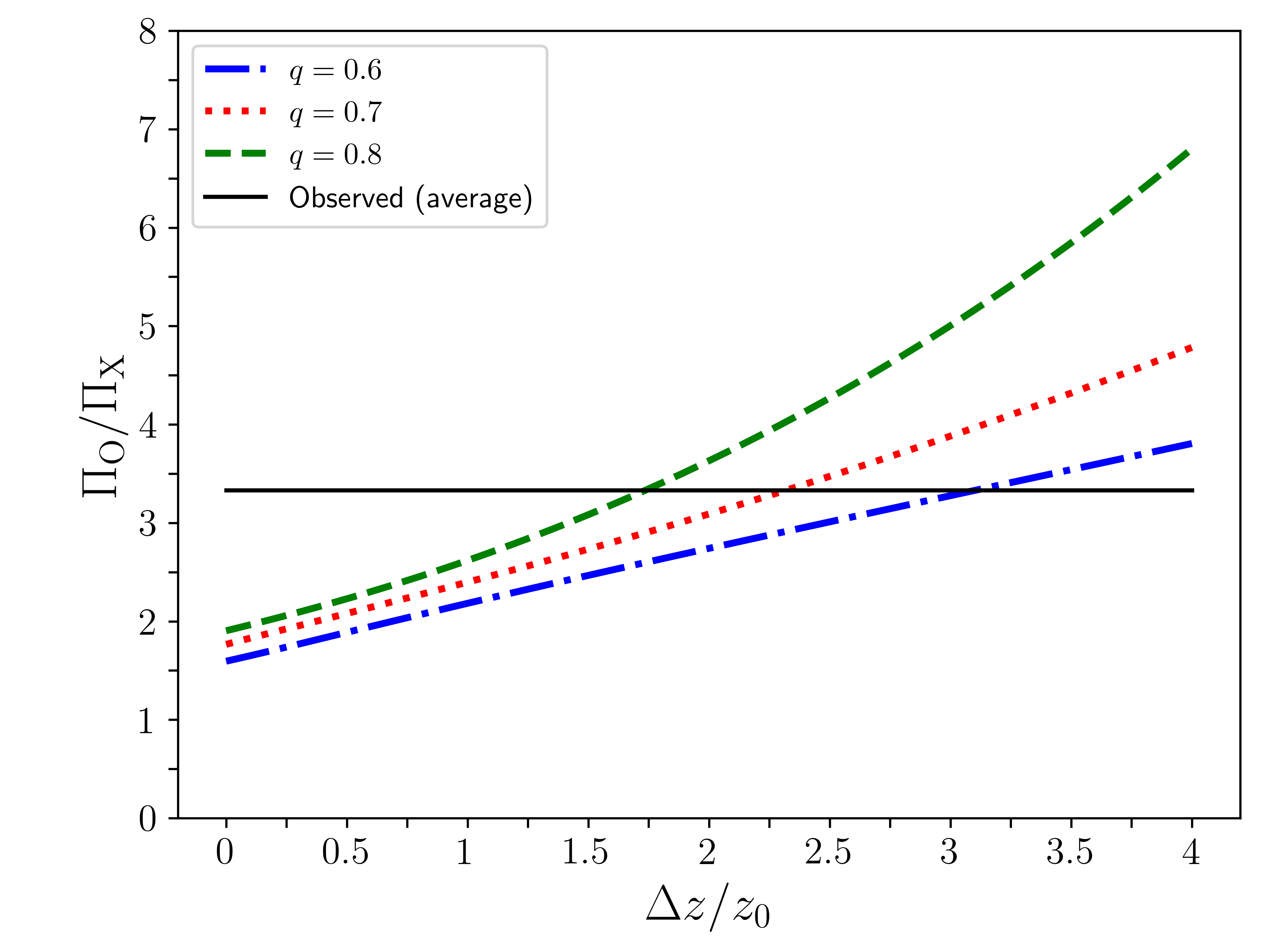}
      \includegraphics[width=0.4\linewidth]{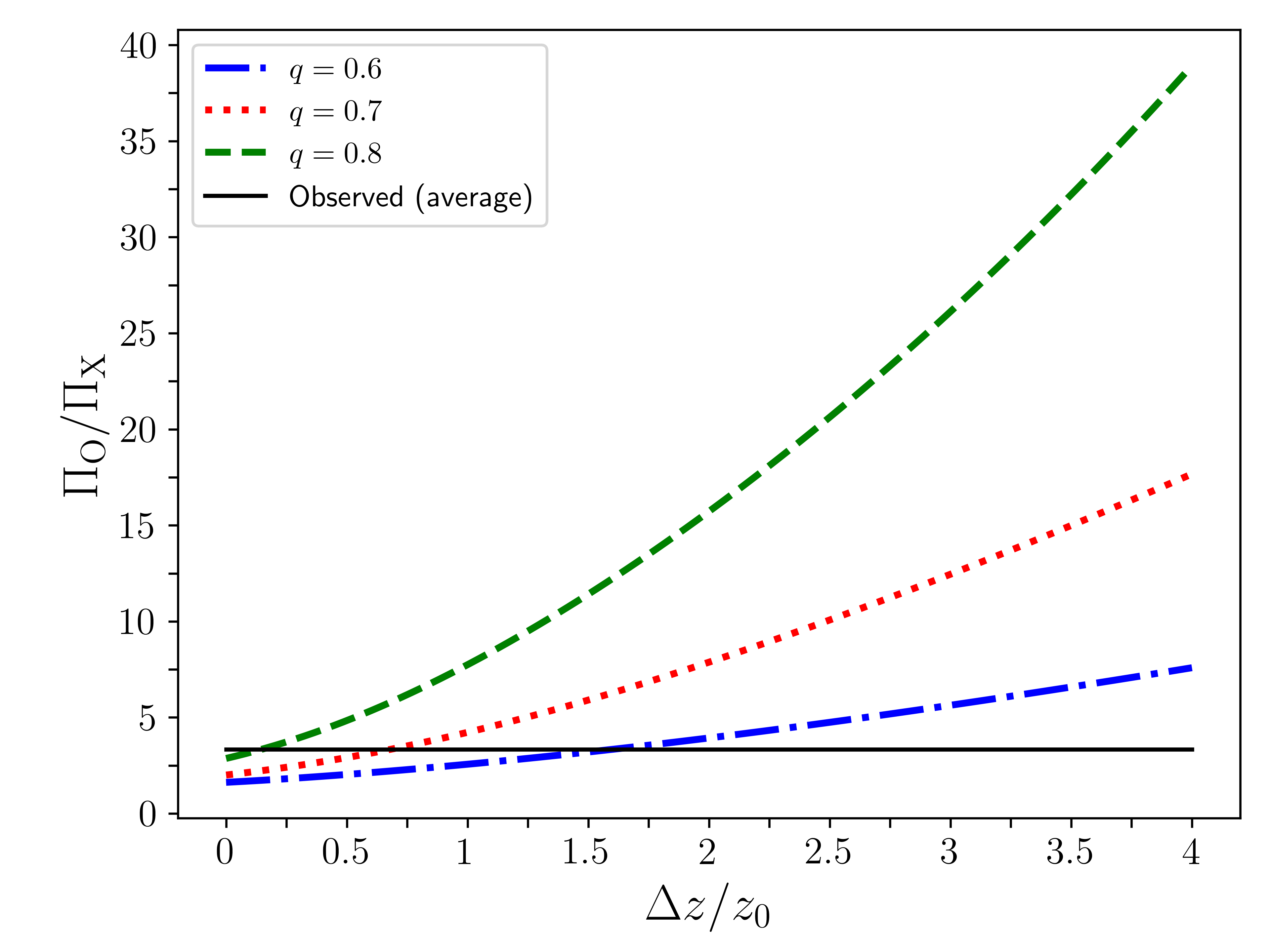}
    \caption{X-ray polarization degree, $\Pi_{\rm{X}}$ (top panels) and chromaticity, $\Pi_{\rm{O}} / \Pi_{\rm{X}}$ (bottom panels) as a function of the width of the proton emission region, $\Delta z$. In the left panels, the proper number density is assumed to be constant, whereas in the right panels, the magnetization is assumed to be constant. The X-ray polarization degree is $\Pi_{\rm X}=\Pi_{p_{\rm p}=2.3}$, and the optical polarization degree is $\Pi_{\rm O}=\Pi_{p_{\rm e}=4.6}$.
    The viewing angle is assumed to be $\theta_{\rm obs}=0.1{\rm\; rad}$.}
    \label{fig:plot}
\end{figure*}

\subsection{Results}

The optical emission is due to synchrotron radiation from nonthermal electrons, whereas we assumed that the X-ray emission is produced through the proton-synchrotron mechanism. As discussed above, in order to calculate the polarization of synchrotron radiation, we assumed that the emitting electrons are confined within an infinitely thin shell (i.e.,~$\Delta z \rightarrow 0$), whereas the protons propagate to large distances, filling a volume with $ 0 < \Delta z < 4 z_0$. We assumed that the electrons emitting in the optical band have a slope $p_{\rm e}=4.6$ and that the protons emitting in the X-ray band have $p_{\rm p}=2.3$ \citep{Agudo25}.

\begin{table}
\caption{\label{table:results}X-ray and optical polarization degree for different parameters.}
\centering
\vskip 0.4cm
\begin{tabular}{l|lcccc}
\toprule
 & $q$ & $\Delta z \rightarrow 0$ & $\Delta z = 2z_0$ & $\Delta z = 4z_0$ \\
\midrule
\multirow{3}{*}{\shortstack[l]{$K_{\rm e,p}=K_0$}} 
    & 0.6 & $\Pi_{\rm O} = 0.449$ & $\Pi_{\rm X} = 0.164$ & $\Pi_{\rm X} = 0.118$ \\
    & 0.7 & $\Pi_{\rm O} = 0.314$ & $\Pi_{\rm X} = 0.101$ & $\Pi_{\rm X} = 0.066$ \\
    & 0.8 & $\Pi_{\rm O} = 0.429$ & $\Pi_{\rm X} = 0.118$ & $\Pi_{\rm X} = 0.063$ \\
\midrule
\multirow{3}{*}{\shortstack[l]{$K_{\rm e,p}\propto B^{\prime 2}$}} 
    & 0.6 & $\Pi_{\rm O} = 0.441$ & $\Pi_{\rm X} = 0.112$ & $\Pi_{\rm X} = 0.058$ \\
    & 0.7 & $\Pi_{\rm O} = 0.256$ & $\Pi_{\rm X} = 0.032$ & $\Pi_{\rm X} = 0.014$ \\
    & 0.8 & $\Pi_{\rm O} = 0.178$ & $\Pi_{\rm X} = 0.011$ & $\Pi_{\rm X} = 0.005$ \\
\bottomrule
\end{tabular}
\vskip 0.4 true cm
\tablefoot{We report $\Pi_{\rm X}$ and $\Pi_{\rm O}$ for different $q$ and different sizes of the particles emission region, $\Delta z$. In the top group, the proper number density is assumed to be constant, whereas in the bottom group, the magnetization is assumed to be constant. The viewing angle is assumed to be $\theta_{\rm obs}=0.1{\rm\; rad}$.}
\vskip -0.4cm
\end{table}

Our results are summarized in Table \ref{table:results} and Fig.~\ref{fig:plot}. In the top panels of Fig.~\ref{fig:plot} we show the observed degree of polarization in the X-ray band as a function of the length of the emission region in units of the distance from the black hole, $\Delta z/z_0$. We consider different jet shapes, described by the parameter $q$. In the left panel, we assume that the density in the proper frame is constant, whereas in the right panel we assume that the magnetization is constant. In the bottom panels, we show the ratio between the degree of polarization in optical and X-rays. The observed degree of polarization (average $\Pi_{\rm O}\sim 25-30\%$ and $\Pi_{\rm X}<7\%$) can be reproduced in both cases (constant number density or constant magnetization) for $q=0.6-0.7$ and $\Delta z\gtrsim 3$. In general, the case of constant number density leads to higher values of $\Pi_{\rm O}$, and the case of constant magnetization leads to smaller $\Pi_{\rm X}$. The electric vector position angle, which is the same for X-ray and optical emission (consistent with the observations), is aligned with the jet axis.

In Appendix \ref{sec:angle} we show the degree of polarization in the X-rays (for a fixed $\Delta z/z_0=3$) and in the optical band as a function of the viewing angle $\theta_{\rm obs}$. Generally, the degree of polarization is proportional to $\theta_{\rm obs}^2$ \citep{Bolis+2024b}. We emphasize that, in the case of constant magnetization, a larger $\Pi_{\rm O}$ can be obtained for $\theta_{\rm obs}\gtrsim 0.1 {\rm\; rad}$, while $\Pi_{\rm X}$ is safely below the IXPE limit.

\section{Discussion}
\label{sec:discussion}

Multifrequency polarimetric observations of BL Lac show different degrees of polarization in the optical band and in the X-rays, which is interpreted as evidence that the origin of the X-rays is leptonic \citep{Agudo25, liodakis25}. We have shown that these observations can also be explained by a hadronic model, where optical radiation is produced through synchrotron emission by relativistic electrons (or, more generally, leptons) and X-rays are produced through the proton-synchrotron mechanism.

The significant difference between the polarization of the synchrotron radiation from the two species is due to the following effects: (i) the different slopes of the underlying particle energy distributions (with slopes $p_{\rm e}=4.6$ and $p_{\rm p}=2.3$ for electrons and protons, respectively), and (ii) the larger emission volume of the protons, which have a longer cooling length. In our model we assumed a globally ordered magnetic field. The introduction of turbulence at intermediate spatial scales would likely further reduce the degree of polarization in the X-ray band \citep{Bandiera2016, Bandiera2024}.

In our scenario, synchrotron emission is the main radiative channel for protons. Photomeson production, considering only the photons internally produced in the jet, provides a negligible contribution\footnote{Correspondingly, the expected neutrino flux is very low. A simple estimate using Eq. 5 of \cite{Murase12} shows that the neutrino flux in the 300 TeV-1 PeV range is on the order of $10^{-13}$ erg cm$^{-2}$ s$^{-1}$, well below the sensitivity of current neutrino detectors.}. However, as commonly found in hadronic scenarios, the radiative cooling times are very long and the proton energy losses are dictated by adiabatic expansion, resulting in a cooling length comparable to the distance of the emitting region from the central black hole. Despite the low radiative efficiency, a recognized problem for hadronic models \citep[e.g.,][]{Zdziarski15}, in our model the jet energy flux required to power the observed emission is below the Eddington luminosity of the associated black hole. 

\citet{liodakis25} explored a hybrid leptonic-hadronic (SSC+hadronic) scenario and a purely hadronic model to interpret the observed variability and polarization signatures of BL Lac. They conclude that their hadronic scenarios cannot reproduce observations due to the long proton cooling time (only radiative losses were considered, and therefore the predicted variability timescale at X-ray/gamma-ray energies is on the order of several years) and the degree of polarization being significantly higher than the IXPE upper limit. Both problems are solved in our model, where the cooling time is dictated by adiabatic losses (which are much more effective than radiative losses) and the polarization is intrinsically low.

In this work we considered a stationary model suitable for reproducing the average polarimetric quantities reported by \cite{Agudo25}. One striking result of the monitoring is the large variability observed in the optical band for both the flux and the degree of polarization (accompanied by moderate activity in the X-ray band). In particular, the degree of polarization displayed a symmetric ``flare'' of about 10 days with variations of 10-45\%. In the framework of our model, this behavior could be reproduced by a small emission region (possibly caused by magnetic reconnection) moving on a helical path with the drift velocity \citep[e.g.,][]{digesu23,Pacciani25}. This scheme naturally predicts an intensity flare accompanied by a large increase in the degree of polarization (approaching values corresponding to a uniform field) when the beamed emission of the moving feature is directed toward the observer. The detailed model will be reported elsewhere (Bolis et al. in prep.).

In this work we focused on the optical and X-ray bands. As is usual for blazar models, the radio data span a spectral region where synchrotron self-absorption is expected to be important. In this particular case, the very hard spectrum reported in Table 4 of \citet{Agudo25} implies that the emission is already partly self-absorbed at frequencies around $10^{12}{\rm\; Hz}$. The radio emission comes from a region of the jet that is located farther downstream with respect to the region where the optical and X-ray emission is produced. Therefore, our scheme does not allow us to estimate the polarimetric properties of radio emission.

In view of future missions with polarimetric capabilities in the MeV (and possibly GeV) bands, for example the Compton Spectrometer and Imager \citep[COSI;][]{tomsick22}, it is useful to look at the expected polarimetric properties in these bands. The high-energy spectrum of BL Lac (and of LSP blazars in general) approximately maintains the same slope up to the peak around $100 {\rm\; MeV}$. Since the cooling length of protons is determined by the adiabatic losses, which are independent of energy, we expect a similar low degree of polarization in the MeV and X-ray bands. On the other hand, after the high-energy SED peak ($E>100{\rm\; MeV}$), the softening of the spectrum determines the increase in the degree of polarization, which could reach 15-20\% in the scenario with constant density.

\begin{acknowledgements}
We thank the referee for useful comments. We acknowledge financial support from an INAF Theory Grant 2022 (PI F.~Tavecchio) and from a Rita Levi Montalcini fellowship (PI E.~Sobacchi). This work has been funded by the European Union-Next Generation EU, PRIN 2022 RFF M4C21.1 (2022C9TNNX). This work has been funded by ASI under contract 2024-11-HH.0.
\end{acknowledgements}

\bibliographystyle{aa}
\bibliography{tavecchio}

\begin{thebibliography}{40}
\expandafter\ifx\csname natexlab\endcsname\relax\def\natexlab#1{#1}\fi

\bibitem[{{Abdo} {et~al.}(2010){Abdo}, {Ackermann}, {Agudo}, {Ajello}, {Aller},
  {Aller}, {Angelakis}, {Arkharov}, {Axelsson}, {Bach}, {Baldini}, {Ballet},
  {Barbiellini}, {Bastieri}, {Baughman}, {Bechtol}, {Bellazzini}, {Benitez},
  {Berdyugin}, {Berenji}, {Blandford}, {Bloom}, {Boettcher}, {Bonamente},
  {Borgland}, {Bregeon}, {Brez}, {Brigida}, {Bruel}, {Burnett}, {Burrows},
  {Buson}, {Caliandro}, {Calzoletti}, {Cameron}, {Capalbi}, {Caraveo},
  {Carosati}, {Casandjian}, {Cavazzuti}, {Cecchi}, {{\c{C}}elik}, {Charles},
  {Chaty}, {Chekhtman}, {Chen}, {Chiang}, {Chincarini}, {Ciprini}, {Claus},
  {Cohen-Tanugi}, {Colafrancesco}, {Cominsky}, {Conrad}, {Costamante},
  {Cutini}, {D'ammando}, {Deitrick}, {D'Elia}, {Dermer}, {de Angelis}, {de
  Palma}, {Digel}, {Donnarumma}, {Silva}, {Drell}, {Dubois}, {Dultzin},
  {Dumora}, {Falcone}, {Farnier}, {Favuzzi}, {Fegan}, {Focke}, {Forn{\'e}},
  {Fortin}, {Frailis}, {Fuhrmann}, {Fukazawa}, {Funk}, {Fusco}, {G{\'o}mez},
  {Gargano}, {Gasparrini}, {Gehrels}, {Germani}, {Giebels}, {Giglietto},
  {Giommi}, {Giordano}, {Giuliani}, {Glanzman}, {Godfrey}, {Grenier},
  {Gronwall}, {Grove}, {Guillemot}, {Guiriec}, {Gurwell}, {Hadasch},
  {Hanabata}, {Harding}, {Hayashida}, {Hays}, {Healey}, {Heidt}, {Hiriart},
  {Horan}, {Hoversten}, {Hughes}, {Itoh}, {Jackson}, {J{\'o}hannesson},
  {Johnson}, {Johnson}, {Jorstad}, {Kadler}, {Kamae}, {Katagiri}, {Kataoka},
  {Kawai}, {Kennea}, {Kerr}, {Kimeridze}, {Kn{\"o}dlseder}, {Kocian},
  {Kopatskaya}, {Koptelova}, {Konstantinova}, {Kovalev}, {Kovalev},
  {Kurtanidze}, {Kuss}, {Lande}, {Larionov}, {Latronico}, {Leto}, {Lindfors},
  {Longo}, {Loparco}, {Lott}, {Lovellette}, {Lubrano}, {Madejski}, {Makeev},
  {Marchegiani}, {Marscher}, {Marshall}, {Max-Moerbeck}, {Mazziotta},
  {McConville}, {McEnery}, {Meurer}, {Michelson}, {Mitthumsiri}, {Mizuno},
  {Moiseev}, {Monte}, {Monzani}, {Morselli}, {Moskalenko}, {Murgia},
  {Nestoras}, {Nilsson}, {Nizhelsky}, {Nolan}, {Norris}, {Nuss}, {Ohsugi},
  {Ojha}, {Omodei}, {Orlando}, {Ormes}, {Osborne}, {Ozaki}, {Pacciani},
  {Padovani}, {Pagani}, {Page}, {Paneque}, {Panetta}, {Parent}, {Pasanen},
  {Pavlidou}, {Pelassa}, {Pepe}, {Perri}, {Pesce-Rollins}, {Piranomonte},
  {Piron}, {Pittori}, {Porter}, {Puccetti}, {Rahoui}, {Rain{\`o}}, {Raiteri},
  {Rando}, {Razzano}, {Reimer}, \& {Reimer}}]{Abdo10}
{Abdo}, A.~A., {Ackermann}, M., {Agudo}, I., {et~al.} 2010, \apj, 716, 30

\bibitem[{{Agudo} {et~al.}(2025){Agudo}, {Liodakis}, {Otero-Santos}, {Middei},
  {Marscher}, {Jorstad}, {Zhang}, {Li}, {Di Gesu}, {Romani}, {Kim}, {Fenu},
  {Marshall}, {Pacciani}, {Escudero Pedrosa}, {Aceituno}, {Agis-Gonzalez},
  {Bonnoli}, {Casanova}, {Morcuende}, {Piirola}, {Sota}, {Kouch}, {Lindfors},
  {McCall}, {Jermak}, {Steele}, {Borman}, {Grishina}, {Hagen-Thorn},
  {Kopatskaya}, {Larionova}, {Morozova}, {Savchenko}, {Shishkina}, {Troitskiy},
  {Troitskaya}, {Vasilyev}, {Zhovtan}, {Myserlis}, {Gurwell}, {Keating}, {Rao},
  {Kang}, {Lee}, {Kim}, {Cheong}, {Jeong}, {Angelakis}, {Kraus}, {Blinov},
  {Maharana}, {Bachev}, {Jormanainen}, {Nilsson}, {Fallah Ramazani}, {Casadio},
  {Fuentes}, {Traianou}, {Thum}, {Gomez}, {Antonelli}, {Bachetti}, {Baldini},
  {Baumgartner}, {Bellazzini}, {Bianchi}, {Bongiorno}, {Bonino}, {Brez},
  {Bucciantini}, {Capitanio}, {Castellano}, {Cavazzuti}, {Chen}, {Ciprini},
  {Costa}, {De Rosa}, {Del Monte}, {Di Lalla}, {Di Marco}, {Donnarumma},
  {Doroshenko}, {Dovciak}, {Ehlert}, {Enoto}, {Evangelista}, {Fabiani},
  {Ferrazzoli}, {Garcia}, {Gunji}, {Hayashida}, {Heyl}, {Iwakiri}, {Kaaret},
  {Karas}, {Kislat}, {Kitaguchi}, {Kolodziejczak}, {Krawczynski}, {La Monaca},
  {Latronico}, {Maldera}, {Manfreda}, {Marin}, {Marinucci}, {Massaro}, {Matt},
  {Mitsuishi}, {Mizuno}, {Muleri}, {Negro}, {Ng}, {O'Dell}, {Omodei},
  {Oppedisano}, {Papitto}, {Pavlov}, {Peirson}, {Perri}, {Pesce-Rollins},
  {Petrucci}, {Pilia}, {Possenti}, {Poutanen}, {Puccetti}, {Ramsey}, {Rankin},
  {Ratheesh}, {Roberts}, {Sgro}, {Slane}, {Soffitta}, {Spandre}, {Swartz},
  {Tamagawa}, {Tavecchio}, {Taverna}, {Tawara}, {Tennant}, {Thomas}, {Tombesi},
  {Trois}, {Tsygankov}, {Turolla}, {Vink}, {Weisskopf}, {Wu}, {Xie}, \&
  {Zane}}]{Agudo25}
{Agudo}, I., {Liodakis}, I., {Otero-Santos}, J., {et~al.} 2025, arXiv e-prints,
  arXiv:2505.01832

\bibitem[{{Aharonian}(2000)}]{Aharonian2000}
{Aharonian}, F.~A. 2000, \na, 5, 377

\bibitem[{{Bandiera} \& {Petruk}(2016)}]{Bandiera2016}
{Bandiera}, R. \& {Petruk}, O. 2016, \mnras, 459, 178

\bibitem[{{Bandiera} \& {Petruk}(2024)}]{Bandiera2024}
{Bandiera}, R. \& {Petruk}, O. 2024, \aap, 689, A137

\bibitem[{{Blandford} {et~al.}(2019){Blandford}, {Meier}, \&
  {Readhead}}]{Blandford19}
{Blandford}, R., {Meier}, D., \& {Readhead}, A. 2019, \araa, 57, 467

\bibitem[{{Bolis} {et~al.}(2024{\natexlab{a}}){Bolis}, {Sobacchi}, \&
  {Tavecchio}}]{Bolis+2024}
{Bolis}, F., {Sobacchi}, E., \& {Tavecchio}, F. 2024{\natexlab{a}}, \aap, 690,
  A14

\bibitem[{{Bolis} {et~al.}(2024{\natexlab{b}}){Bolis}, {Sobacchi}, \&
  {Tavecchio}}]{Bolis+2024b}
{Bolis}, F., {Sobacchi}, E., \& {Tavecchio}, F. 2024{\natexlab{b}}, \prd, 110,
  123032

\bibitem[{{Bonometto} \& {Saggion}(1973)}]{Bonometto73}
{Bonometto}, S. \& {Saggion}, A. 1973, \aap, 23, 9

\bibitem[{{B{\"o}ttcher}(2019)}]{Boettcher19}
{B{\"o}ttcher}, M. 2019, Galaxies, 7, 20

\bibitem[{{Capetti} {et~al.}(2010){Capetti}, {Raiteri}, \&
  {Buttiglione}}]{Capetti10}
{Capetti}, A., {Raiteri}, C.~M., \& {Buttiglione}, S. 2010, \aap, 516, A59

\bibitem[{{Del Zanna} {et~al.}(2006){Del Zanna}, {Volpi}, {Amato}, \&
  {Bucciantini}}]{DelZanna06}
{Del Zanna}, L., {Volpi}, D., {Amato}, E., \& {Bucciantini}, N. 2006, \aap,
  453, 621

\bibitem[{{Di Gesu} {et~al.}(2023){Di Gesu}, {Marshall}, {Ehlert}, {Kim},
  {Donnarumma}, {Tavecchio}, {Liodakis}, {Kiehlmann}, {Agudo}, {Jorstad},
  {Muleri}, {Marscher}, {Puccetti}, {Middei}, {Perri}, {Pacciani}, {Negro},
  {Romani}, {Di Marco}, {Blinov}, {Bourbah}, {Kontopodis}, {Mandarakas},
  {Romanopoulos}, {Skalidis}, {Vervelaki}, {Casadio}, {Escudero}, {Myserlis},
  {Gurwell}, {Rao}, {Keating}, {Kouch}, {Lindfors}, {Aceituno}, {Bernardos},
  {Bonnoli}, {Casanova}, {Garc{\'\i}a-Comas}, {Ag{\'\i}s-Gonz{\'a}lez},
  {Husillos}, {Marchini}, {Sota}, {Imazawa}, {Sasada}, {Fukazawa}, {Kawabata},
  {Uemura}, {Mizuno}, {Nakaoka}, {Akitaya}, {Savchenko}, {Vasilyev},
  {G{\'o}mez}, {Antonelli}, {Barnouin}, {Bonino}, {Cavazzuti}, {Costamante},
  {Chen}, {Cibrario}, {De Rosa}, {Di Pierro}, {Errando}, {Kaaret}, {Karas},
  {Krawczynski}, {Lisalda}, {Madejski}, {Malacaria}, {Marin}, {Marinucci},
  {Massaro}, {Matt}, {Mitsuishi}, {O'Dell}, {Paggi}, {Peirson}, {Petrucci},
  {Ramsey}, {Tennant}, {Wu}, {Bachetti}, {Baldini}, {Baumgartner},
  {Bellazzini}, {Bianchi}, {Bongiorno}, {Brez}, {Bucciantini}, {Capitanio},
  {Castellano}, {Ciprini}, {Costa}, {Del Monte}, {Di Lalla}, {Doroshenko},
  {Dov{\v{c}}iak}, {Enoto}, {Evangelista}, {Fabiani}, {Ferrazzoli}, {Garcia},
  {Gunji}, {Hayashida}, {Heyl}, {Iwakiri}, {Kislat}, {Kitaguchi},
  {Kolodziejczak}, {La Monaca}, {Latronico}, {Maldera}, {Manfreda}, {Ng},
  {Omodei}, {Oppedisano}, {Papitto}, {Pavlov}, {Pesce-Rollins}, {Pilia},
  {Possenti}, {Poutanen}, {Rankin}, {Ratheesh}, {Roberts}, {Sgr{\`o}}, {Slane},
  {Soffitta}, {Spandre}, {Swartz}, {Tamagawa}, {Taverna}, {Tawara}, {Thomas},
  {Tombesi}, {Trois}, {Tsygankov}, {Turolla}, {Vink}, {Weisskopf}, {Xie}, \&
  {Zane}}]{digesu23}
{Di Gesu}, L., {Marshall}, H.~L., {Ehlert}, S.~R., {et~al.} 2023, Nature
  Astronomy, 7, 1245

\bibitem[{{Fossati} {et~al.}(1998){Fossati}, {Maraschi}, {Celotti}, {Comastri},
  \& {Ghisellini}}]{fossati98}
{Fossati}, G., {Maraschi}, L., {Celotti}, A., {Comastri}, A., \& {Ghisellini},
  G. 1998, \mnras, 299, 433

\bibitem[{{Ghisellini} {et~al.}(1998){Ghisellini}, {Celotti}, {Fossati},
  {Maraschi}, \& {Comastri}}]{ghisellini98}
{Ghisellini}, G., {Celotti}, A., {Fossati}, G., {Maraschi}, L., \& {Comastri},
  A. 1998, \mnras, 301, 451

\bibitem[{{Hillas}(1984)}]{Hillas}
{Hillas}, A.~M. 1984, \araa, 22, 425

\bibitem[{{Krawczynski}(2012)}]{Krawczynski12}
{Krawczynski}, H. 2012, \apj, 744, 30

\bibitem[{{Liodakis} \& {Petropoulou}(2020)}]{liodakis20}
{Liodakis}, I. \& {Petropoulou}, M. 2020, \apjl, 893, L20

\bibitem[{{Liodakis} {et~al.}(2025){Liodakis}, {Zhang}, {Boula}, {Middei},
  {Otero-Santos}, {Blinov}, {Agudo}, {B{\"o}ttcher}, {Chen}, {Ehlert},
  {Jorstad}, {Kaaret}, {Krawczynski}, {Peirson}, {Romani}, {Tavecchio},
  {Weisskopf}, {Kouch}, {Lindfors}, {Nilsson}, {McCall}, {Jermak}, {Steele},
  {Myserlis}, {Gurwell}, {Keating}, {Rao}, {Kang}, {Lee}, {Kim}, {Cheong},
  {Jeong}, {Angelakis}, {Kraus}, {Jos{\'e} Aceituno}, {Bonnoli}, {Casanova},
  {Escudero}, {Ag{\'\i}s-Gonz{\'a}lez}, {Morcuende}, {Sota}, {Bachev},
  {Grishina}, {Kopatskaya}, {Larionova}, {Morozova}, {Savchenko}, {Shishkina},
  {Troitskiy}, {Troitskaya}, \& {Vasilyev}}]{liodakis25}
{Liodakis}, I., {Zhang}, H., {Boula}, S., {et~al.} 2025, arXiv e-prints,
  arXiv:2505.13603

\bibitem[{{Lyubarsky}(2009)}]{Lyubarsky2009}
{Lyubarsky}, Y. 2009, \apj, 698, 1570

\bibitem[{{Lyutikov} {et~al.}(2003){Lyutikov}, {Pariev}, \&
  {Blandford}}]{Lyutikov2003}
{Lyutikov}, M., {Pariev}, V.~I., \& {Blandford}, R.~D. 2003, \apj, 597, 998

\bibitem[{{Marshall} {et~al.}(2024){Marshall}, {Liodakis}, {Marscher}, {Di
  Lalla}, {Jorstad}, {Kim}, {Middei}, {Negro}, {Omodei}, {Peirson}, {Perri},
  {Puccetti}, {Laurenti}, {Agudo}, {Bonnoli}, {Berdyugin}, {Cavazzuti},
  {Rodriguez Cavero}, {Donnarumma}, {Di Gesu}, {Jormanainen}, {Krawczynski},
  {Lindfors}, {Madjeski}, {Marin}, {Massaro}, {Pacciani}, {Poutanen},
  {Tavecchio}, {Kouch}, {Aceituno}, {Bernardos}, {Casanova},
  {Garc{\'\i}a-Comas}, {Ag{\'\i}s-Gonz{\'a}lez}, {Husillos}, {Marchini},
  {Sota}, {Blinov}, {Bourbah}, {Kielhmann}, {Kontopodis}, {Mandarakas},
  {Romanopoulos}, {Skalidis}, {Vervelaki}, {Borman}, {Kopatskaya}, {Larionova},
  {Morozova}, {Savchenko}, {Vasilyev}, {Zhovtan}, {Casadio}, {Escudero},
  {Kramer}, {Myserlis}, {Trainou}, {Imazawa}, {Sasada}, {Fukazawa}, {Kawabata},
  {Uemura}, {Mizuno}, {Nakaoka}, {Akitaya}, {Masiero}, {Mawet}, {Panopoulou},
  {Tinyanont}, {Kagitani}, {Kravtsov}, {Sakanoi}, {Dattolo}, {Gurwell},
  {Keating}, {Rao}, {Cheong}, {Jeong}, {Kang}, {Kim}, {Lee}, {Angelakis},
  {Kraus}, {Hales}, {Kameno}, {Kneissl}, {Messias}, {Nagai}, {Antonelli},
  {Bachetti}, {Baldini}, {Baumgartner}, {Bellazzini}, {Bianchi}, {Bongiorno},
  {Bonino}, {Brez}, {Bucciantini}, {Capitanio}, {Castellano}, {Chen},
  {Ciprini}, {Costa}, {De Rosa}, {Del Monte}, {Di Marco}, {Doroshenko},
  {Dov{\v{c}}iak}, {Ehlert}, {Enoto}, {Evangelista}, {Fabiani}, {Ferrazzoli},
  {Garcia}, {Gunji}, {Hayashida}, {Heyl}, {Iwakiri}, {Kaaret}, {Karas},
  {Kislat}, {Kitaguchi}, {Kolodziejczak}, {La Monaca}, {Latronico}, {Maldera},
  {Manfreda}, {Marinucci}, {Matt}, {Mitsuishi}, {Muleri}, {Ng}, {O'Dell},
  {Oppedisano}, {Papitto}, {Pavlov}, {Pesce-Rollins}, {Petrucci}, {Pilia},
  {Possenti}, {Ramsey}, {Rankin}, {Ratheesh}, {Roberts}, {Romani}, {Sgr{\`o}},
  {Slane}, {Soffitta}, {Spandre}, {Swartz}, {Tamagawa}, {Taverna}, {Tawara},
  {Tennant}, {Thomas}, {Tombesi}, {Trois}, {Tsygankov}, {Turolla}, {Vink},
  {Weisskopf}, {Wu}, {Xie}, \& {Zane}}]{Marshall24}
{Marshall}, H.~L., {Liodakis}, I., {Marscher}, A.~P., {et~al.} 2024, \apj, 972,
  74

\bibitem[{{Middei} {et~al.}(2023){Middei}, {Liodakis}, {Perri}, {Puccetti},
  {Cavazzuti}, {Di Gesu}, {Ehlert}, {Madejski}, {Marscher}, {Marshall},
  {Muleri}, {Negro}, {Jorstad}, {Ag{\'\i}s-Gonz{\'a}lez}, {Agudo}, {Bonnoli},
  {Bernardos}, {Casanova}, {Garc{\'\i}a-Comas}, {Husillos}, {Marchini}, {Sota},
  {Kouch}, {Lindfors}, {Borman}, {Kopatskaya}, {Larionova}, {Morozova},
  {Savchenko}, {Vasilyev}, {Zhovtan}, {Casadio}, {Escudero}, {Myserlis},
  {Hales}, {Kameno}, {Kneissl}, {Messias}, {Nagai}, {Blinov}, {Bourbah},
  {Kiehlmann}, {Kontopodis}, {Mandarakas}, {Romanopoulos}, {Skalidis},
  {Vervelaki}, {Masiero}, {Mawet}, {Millar-Blanchaer}, {Panopoulou},
  {Tinyanont}, {Berdyugin}, {Kagitani}, {Kravtsov}, {Sakanoi}, {Imazawa},
  {Sasada}, {Fukazawa}, {Kawabata}, {Uemura}, {Mizuno}, {Nakaoka}, {Akitaya},
  {Gurwell}, {Rao}, {Di Lalla}, {Cibrario}, {Donnarumma}, {Kim}, {Omodei},
  {Pacciani}, {Poutanen}, {Tavecchio}, {Antonelli}, {Bachetti}, {Baldini},
  {Baumgartner}, {Bellazzini}, {Bianchi}, {Bongiorno}, {Bonino}, {Brez},
  {Bucciantini}, {Capitanio}, {Castellano}, {Ciprini}, {Costa}, {De Rosa}, {Del
  Monte}, {Di Marco}, {Doroshenko}, {Dov{\v{c}}iak}, {Enoto}, {Evangelista},
  {Fabiani}, {Ferrazzoli}, {Garcia}, {Gunji}, {Hayashida}, {Heyl}, {Iwakiri},
  {Karas}, {Kitaguchi}, {Kolodziejczak}, {Krawczynski}, {La Monaca},
  {Latronico}, {Maldera}, {Manfreda}, {Marin}, {Marinucci}, {Massaro}, {Matt},
  {Mitsuishi}, {Ng}, {O'Dell}, {Oppedisano}, {Papitto}, {Pavlov}, {Peirson},
  {Pesce-Rollins}, {Petrucci}, {Pilia}, {Possenti}, {Ramsey}, {Rankin},
  {Ratheesh}, {Romani}, {Sgr{\'o}}, {Slane}, {Soffitta}, {Spandre}, {Tamagawa},
  {Taverna}, {Tawara}, {Tennant}, {Thomas}, {Tombesi}, {Trois}, {Tsygankov},
  {Turolla}, {Vink}, {Weisskopf}, {Wu}, {Xie}, \& {Zane}}]{middei23b}
{Middei}, R., {Liodakis}, I., {Perri}, M., {et~al.} 2023, \apjl, 942, L10

\bibitem[{{Murase} {et~al.}(2012){Murase}, {Dermer}, {Takami}, \&
  {Migliori}}]{Murase12}
{Murase}, K., {Dermer}, C.~D., {Takami}, H., \& {Migliori}, G. 2012, \apj, 749,
  63

\bibitem[{{Pacciani} {et~al.}(2025){Pacciani}, {Kim}, {Middei}, {Marshall},
  {Marscher}, {Liodakis}, {Agudo}, {Jorstad}, {Poutanen}, {Errando}, {Di Gesu},
  {Negro}, {Tavecchio}, {Wu}, {Chen}, {Muleri}, {Antonelli}, {Donnarumma},
  {Ehlert}, {Massaro}, {O'Dell}, {Perri}, {Puccetti}, {Aceituno}, {Bonnoli},
  {Casanova}, {Escudero}, {Ag{\'\i}s-Gonz{\'a}lez}, {Husillos}, {Morcuende},
  {Otero-Santos}, {Sota}, {Kouch}, {Lindfors}, {Borman}, {G{\'o}mez},
  {Kopatskaya}, {Larionova}, {Morozova}, {Savchenko}, {Vasilyev}, {Zhovtan},
  {Blinov}, {Gourni}, {Kiehlmann}, {Kourtidis}, {Mandarakas}, {Palaiologou},
  {Triantafyllou}, {Vervelaki}, {Myserlis}, {Gurwell}, {Keating}, {Rao},
  {Angelakis}, {Kraus}, {Bachetti}, {Baldini}, {Baumgartner}, {Bellazzini},
  {Bianchi}, {Bongiorno}, {Bonino}, {Brez}, {Bucciantini}, {Capitanio},
  {Castellano}, {Cavazzuti}, {Ciprini}, {Costa}, {De Rosa}, {Del Monte}, {Di
  Lalla}, {Di Marco}, {Doroshenko}, {Dov{\v{c}}iak}, {Enoto}, {Evangelista},
  {Fabiani}, {Ferrazzoli}, {Garcia}, {Gunji}, {Hayashida}, {Heyl}, {Iwakiri},
  {Kaaret}, {Karas}, {Kislat}, {Kitaguchi}, {Kolodziejczak}, {Krawczynski}, {La
  Monaca}, {Latronico}, {Maldera}, {Manfreda}, {Marin}, {Marinucci}, {Matt},
  {Mitsuishi}, {Mizuno}, {Ng}, {Omodei}, {Oppedisano}, {Papitto}, {Pavlov},
  {Peirson}, {Pesce-Rollins}, {Petrucci}, {Pilia}, {Possenti}, {Ramsey},
  {Rankin}, {Ratheesh}, {Roberts}, {Romani}, {Sgr{\'o}}, {Slane}, {Soffitta},
  {Spandre}, {Swartz}, {Tamagawa}, {Taverna}, {Tawara}, {Tennant}, {Thomas},
  {Tombesi}, {Trois}, {Tsygankov}, {Turolla}, {Vink}, {Weisskopf}, {Xie}, \&
  {Zane}}]{Pacciani25}
{Pacciani}, L., {Kim}, D.~E., {Middei}, R., {et~al.} 2025, \apj, 983, 78

\bibitem[{{Peirson} {et~al.}(2022){Peirson}, {Liodakis}, \&
  {Romani}}]{peirson22}
{Peirson}, A.~L., {Liodakis}, I., \& {Romani}, R.~W. 2022, \apj, 931, 59

\bibitem[{{Peirson} {et~al.}(2023){Peirson}, {Negro}, {Liodakis}, {Middei},
  {Kim}, {Marscher}, {Marshall}, {Pacciani}, {Romani}, {Wu}, {Di Marco}, {Di
  Lalla}, {Omodei}, {Jorstad}, {Agudo}, {Kouch}, {Lindfors}, {Aceituno},
  {Bernardos}, {Bonnoli}, {Casanova}, {Garc{\'\i}a-Comas},
  {Ag{\'\i}s-Gonz{\'a}lez}, {Husillos}, {Marchini}, {Sota}, {Casadio},
  {Escudero}, {Myserlis}, {Sievers}, {Gurwell}, {Rao}, {Imazawa}, {Sasada},
  {Fukazawa}, {Kawabata}, {Uemura}, {Mizuno}, {Nakaoka}, {Akitaya}, {Cheong},
  {Jeong}, {Kang}, {Kim}, {Lee}, {Angelakis}, {Kraus}, {Cibrario},
  {Donnarumma}, {Poutanen}, {Tavecchio}, {Antonelli}, {Bachetti}, {Baldini},
  {Baumgartner}, {Bellazzini}, {Bianchi}, {Bongiorno}, {Bonino}, {Brez},
  {Bucciantini}, {Capitanio}, {Castellano}, {Cavazzuti}, {Chen}, {Ciprini},
  {Costa}, {De Rosa}, {Del Monte}, {Di Gesu}, {Doroshenko}, {Dov{\v{c}}iak},
  {Ehlert}, {Enoto}, {Evangelista}, {Fabiani}, {Ferrazzoli}, {Garcia}, {Gunji},
  {Hayashida}, {Heyl}, {Iwakiri}, {Kaaret}, {Karas}, {Kitaguchi},
  {Kolodziejczak}, {Krawczynski}, {La Monaca}, {Latronico}, {Madejski},
  {Maldera}, {Manfreda}, {Marin}, {Marinucci}, {Massaro}, {Matt}, {Mitsuishi},
  {Muleri}, {Ng}, {O'Dell}, {Oppedisano}, {Papitto}, {Pavlov}, {Perri},
  {Pesce-Rollins}, {Petrucci}, {Pilia}, {Possenti}, {Puccetti}, {Ramsey},
  {Rankin}, {Ratheesh}, {Roberts}, {Sgr{\'o}}, {Slane}, {Soffitta}, {Spandre},
  {Swartz}, {Tamagawa}, {Taverna}, {Tawara}, {Tennant}, {Thomas}, {Tombesi},
  {Trois}, {Tsygankov}, {Turolla}, {Vink}, {Weisskopf}, {Xie}, \&
  {Zane}}]{Peirson23}
{Peirson}, A.~L., {Negro}, M., {Liodakis}, I., {et~al.} 2023, \apjl, 948, L25

\bibitem[{{Poutanen}(1994)}]{Poutanen94}
{Poutanen}, J. 1994, \apjs, 92, 607

\bibitem[{{Romero} {et~al.}(2017){Romero}, {Boettcher}, {Markoff}, \&
  {Tavecchio}}]{romero17}
{Romero}, G.~E., {Boettcher}, M., {Markoff}, S., \& {Tavecchio}, F. 2017, \ssr,
  207, 5

\bibitem[{{Sikora} {et~al.}(1994){Sikora}, {Begelman}, \& {Rees}}]{sikora94}
{Sikora}, M., {Begelman}, M.~C., \& {Rees}, M.~J. 1994, \apj, 421, 153

\bibitem[{{Sol} \& {Zech}(2022)}]{SolZech22}
{Sol}, H. \& {Zech}, A. 2022, Galaxies, 10, 105

\bibitem[{{Tavecchio} {et~al.}(2003){Tavecchio}, {Ghisellini}, \&
  {Celotti}}]{Tavecchio03}
{Tavecchio}, F., {Ghisellini}, G., \& {Celotti}, A. 2003, \aap, 403, 83

\bibitem[{{Tomsick} \& {COSI Collaboration}(2022)}]{tomsick22}
{Tomsick}, J. \& {COSI Collaboration}. 2022, in 37th International Cosmic Ray
  Conference, 652

\bibitem[{{Vlahakis}(2004)}]{Vlahakis2004}
{Vlahakis}, N. 2004, \apj, 600, 324

\bibitem[{{Webb}(1989)}]{Webb89}
{Webb}, G.~M. 1989, \apj, 340, 1112

\bibitem[{{Weisskopf} {et~al.}(2022){Weisskopf}, {Soffitta}, {Baldini},
  {Ramsey}, {O'Dell}, {Romani}, {Matt}, {Deininger}, {Baumgartner},
  {Bellazzini}, {Costa}, {Kolodziejczak}, {Latronico}, {Marshall}, {Muleri},
  {Bongiorno}, {Tennant}, {Bucciantini}, {Dovciak}, {Marin}, {Marscher},
  {Poutanen}, {Slane}, {Turolla}, {Kalinowski}, {Di Marco}, {Fabiani},
  {Minuti}, {La Monaca}, {Pinchera}, {Rankin}, {Sgro'}, {Trois}, {Xie},
  {Alexander}, {Allen}, {Amici}, {Andersen}, {Antonelli}, {Antoniak},
  {Attin{\`a}}, {Barbanera}, {Bachetti}, {Baggett}, {Bladt}, {Brez}, {Bonino},
  {Boree}, {Borotto}, {Breeding}, {Brienza}, {Bygott}, {Caporale}, {Cardelli},
  {Carpentiero}, {Castellano}, {Castronuovo}, {Cavalli}, {Cavazzuti},
  {Ceccanti}, {Centrone}, {Citraro}, {D'Amico}, {D'Alba}, {Di Gesu}, {Del
  Monte}, {Dietz}, {Di Lalla}, {Persio}, {Dolan}, {Donnarumma}, {Evangelista},
  {Ferrant}, {Ferrazzoli}, {Ferrie}, {Footdale}, {Forsyth}, {Foster},
  {Garelick}, {Gunji}, {Gurnee}, {Head}, {Hibbard}, {Johnson}, {Kelly},
  {Kilaru}, {Lefevre}, {Roy}, {Loffredo}, {Lorenzi}, {Lucchesi}, {Maddox},
  {Magazzu}, {Maldera}, {Manfreda}, {Mangraviti}, {Marengo}, {Marrocchesi},
  {Massaro}, {Mauger}, {McCracken}, {McEachen}, {Mize}, {Mereu}, {Mitchell},
  {Mitsuishi}, {Morbidini}, {Mosti}, {Nasimi}, {Negri}, {Negro}, {Nguyen},
  {Nitschke}, {Nuti}, {Onizuka}, {Oppedisano}, {Orsini}, {Osborne}, {Pacheco},
  {Paggi}, {Painter}, {Pavelitz}, {Pentz}, {Piazzolla}, {Perri},
  {Pesce-Rollins}, {Peterson}, {Pilia}, {Profeti}, {Puccetti}, {Ranganathan},
  {Ratheesh}, {Reedy}, {Root}, {Rubini}, {Ruswick}, {Sanchez}, {Sarra},
  {Santoli}, {Scalise}, {Sciortino}, {Schroeder}, {Seek}, {Sosdian}, {Spandre},
  {Speegle}, {Tamagawa}, {Tardiola}, {Tobia}, {Thomas}, {Valerie}, {Vimercati},
  {Walden}, {Weddendorf}, {Wedmore}, {Welch}, {Zanetti}, \&
  {Zanetti}}]{weisskopf22}
{Weisskopf}, M.~C., {Soffitta}, P., {Baldini}, L., {et~al.} 2022, Journal of
  Astronomical Telescopes, Instruments, and Systems, 8, 026002

\bibitem[{{Zdziarski} \& {Bottcher}(2015)}]{Zdziarski15}
{Zdziarski}, A.~A. \& {Bottcher}, M. 2015, \mnras, 450, L21

\bibitem[{{Zech} {et~al.}(2017){Zech}, {Cerruti}, \& {Mazin}}]{Zech17}
{Zech}, A., {Cerruti}, M., \& {Mazin}, D. 2017, \aap, 602, A25

\bibitem[{{Zhang}(2017)}]{Zhang17}
{Zhang}, H. 2017, Galaxies, 5, 32

\bibitem[{{Zhang} \& {B{\"o}ttcher}(2013)}]{ZhangBoettcher13}
{Zhang}, H. \& {B{\"o}ttcher}, M. 2013, \apj, 774, 18

\end{thebibliography}

\appendix

\section{Limit on SSC emission}
\label{sec:SSC}

For a source with observed synchrotron luminosity $L_{\rm s}$ (for BL Lac, one has $L_{\rm s} \sim 10^{45}{\rm\; erg\; s}^{-1}$), the luminosity of the SSC emission can be estimated as\begin{equation}
L_{\rm SSC}=\frac{U^{\prime}_{\rm s}}{U^{\prime}_{\rm B}} L_{\rm s}\;.
\end{equation}
Here $U^{\prime}_{\rm s}$ is the energy density of the synchrotron radiation in the source frame:
\begin{equation}
    U^{\prime}_{\rm s}=\frac{L_{\rm s}}{4\pi R^2 c \delta^4}\;,
\end{equation}
where $R$ is the source size, and $U^{\prime}_{\rm B}=B^{\prime 2}/8\pi$ is the magnetic field energy density in the source frame. Assuming $B^{\prime}=10{\rm\; G}$, $R\simeq 3\times 10^{16}{\rm\; cm}$, $\delta=10$, we find $L_{\rm SSC}\simeq 5\times 10^{41}{\rm\; erg\; s}^{-1}$, a negligible fraction of the observed luminosity in the high-energy band (X-rays and gamma-rays).

\section{Adiabatic cooling time}
\label{sec:adiabatic}

The general expression for the adiabatic cooling time of relativistic particles in an expanding flow, in the frame of the fluid, is \citep[e.g.,][]{Webb89}
\begin{equation}
    t^{\prime}_{\rm adiab}=\frac{3}{\grad_{\mu} u^{\mu}}\;,
    \label{eq:webb}
\end{equation}
where $\grad_{\mu} u^{\mu}$ is the four-divergence of the fluid four-velocity. Assuming that the flow is stationary and axisymmetric, in cylindrical coordinates the four-divergence can then be written as
\begin{equation}
\grad_{\mu} u^{\mu} = \frac{1}{R} \frac{\partial (\Gamma Rv_R)}{\partial R} + \frac{\partial (\Gamma v_z)}{\partial z}\;.
\label{eq:quadri}
\end{equation}
The radial expansion speed can be approximated as $v_R\simeq c \Theta(R)$, where $\Theta(R)$ is the angle between the jet axis and the flow speed at radius $R$, given by Eq.~\eqref{eq:angle}. 

In the radial derivative, we assumed that $\Gamma $ is approximately constant and use $\Omega R_{0} = 3^{1/4} ( \Omega z_0)^{q}$ (see Sect. \ref{sec:polariz}). Thus, we obtain\begin{equation}
\frac{1}{R} \frac{\partial (\Gamma Rv_R)}{\partial R}\approx \frac{2cq\Gamma}{z}\;.
\end{equation}
For the derivative along $z$ we can write
\begin{equation}
\label{eq:zderiv}
\frac{\partial (\Gamma v_z)}{\partial z}\simeq c \frac{\partial \Gamma}{\partial z}\;.
\end{equation}
Using Eq.~(29) of \cite{Bolis+2024b} to calculate $\Gamma$, Eq.~\eqref{eq:zderiv} can be approximated as
\begin{equation}
    c \frac{\partial \Gamma}{\partial z}\simeq A(q) \Omega \left( \frac{\Omega}{c} z \right)^{-q}\;,
\end{equation}
where $A(q)$ is a factor of order unity. 
One can check that for values of $R$ and $z$ suitable for blazars ($R=3\times 10^{16}{\rm\; cm}$, $z=10^{17}{\rm\; cm}$) this term can be neglected. Therefore, the adiabatic cooling time is given by
\begin{equation}
    t^{\prime}_{\rm adiab}=\frac{3z}{2qc\Gamma}\;.
\end{equation}
The corresponding cooling length in the observer frame is\begin{equation}
    l_{\rm adiab}=ct^{\prime}_{\rm adiab}\Gamma =\frac{3}{2} \frac{z}{q}\;.
\end{equation}
For $q=1$, we recover the standard expression for a conical jet \citep[e.g.,][]{Tavecchio03}.

\section{Jet energy flux}
\label{sec:lp}

The jet energy flux associated with protons can be estimated as follows. The total synchrotron luminosity produced by protons can be written as
\begin{equation}
L_{\rm s,p}\simeq\frac{4}{3} \sigma_{\rm T} c \left( \frac {m_{\rm e}}{m_{\rm p}} \right)^2 \frac{B^{\prime 2}}{8\pi} \mathcal{V} \delta^4 \int n_{\rm p}(\gamma_{\rm p})\gamma_{\rm p}^2 d\gamma_{\rm p}\;,
\label{eq:lsprot}
\end{equation}
\noindent
where $n_{\rm p}(\gamma_{\rm p})$ is the proton energy distribution and the volume is $\mathcal{V}\simeq \pi R_0^2 \Delta z$. 

Assuming $n_{\rm p}(\gamma_{\rm p})=k \gamma_{\rm p}^{-2}$ in the range $\gamma_{\rm p,1}<\gamma _{\rm p} < \gamma _{\rm p,2}$, the integral in Eq.~\eqref{eq:lsprot} is
\begin{equation}
    \int_{\gamma_{\rm p,1}}^{\gamma_{\rm p,2}} n_{\rm p}(\gamma_{\rm p})\gamma_{\rm p}^2 d\gamma_{\rm p} \simeq k \, \gamma_{\rm p,2}\;.
\end{equation}
It is now possible to express the normalization of the energy distribution as 
\begin{equation}
k = \frac{6\pi \, L_{\rm s,p}}{\sigma_{\rm T} c \,B^{\prime 2} \, \gamma_{\rm p,2} \delta^4 \,\mathcal{V} } \left( \frac {m_{\rm p}}{m_{\rm e}} \right)^2\;.
\label{eq:k}
\end{equation}
The energy flux carried by protons is\begin{equation}
    P_{\rm p} =\pi R_0^2 U^{\prime}_{\rm p} \Gamma ^2 c\;,
    \label{eq:pp}
\end{equation}
where the proton energy density is 
\begin{equation}
    U^{\prime}_{\rm p} = m_{\rm p} c^2 \int_{\gamma_{\rm p,1}}^{\gamma_{\rm p,2}} n_{\rm p}(\gamma_{\rm p})\gamma_{\rm p} d\gamma_{\rm p} =  k\, m_{\rm p} c^2 \ln \frac{\gamma_{\rm p,2}}{\gamma_{\rm p,1}}\;.
    \label{eq:up}
\end{equation}
Inserting Eq.~\eqref{eq:up} in Eq.~\eqref{eq:pp} and using Eq.~\eqref{eq:k}, one can calculate $P_{\rm p}$. Assuming $B^{\prime}=10{\rm\; G}$, $\Delta z=3\times 10^{17}{\rm\; cm}$, $\gamma_{\rm p,1}=1$ $\gamma_{\rm p,2}=3\times 10^8$, $\delta=10$, $L_{\rm s,p} \sim 10^{45}{\rm\; erg\; s}^{-1}$, we derive $P_{\rm p}=3\times 10^{45}{\rm\; erg\; s}^{-1}$.

\section{Stokes parameters}
\label{sec:Stokes}

The Stokes parameters of a stationary jet can be presented as \citep{Lyutikov2003, DelZanna06}
\begin{align}
\label{eq:stokes1}
I & = \frac{p + 7/3}{p+1} \; \kappa_{\rm s} \int {\rm d}V \; K_{\rm e,p} \; \delta^{\left( 3 + p\right)/2} \;  \left| \mathbf{B}' \times \hat{\mathbf{n}}' \right|^{\left( p+1 \right)/2} \\
\label{eq:stokes2}
Q & = \kappa_{\rm s} \int {\rm d}V \; K_{\rm e,p} \; \delta^{\left( 3 + p\right)/2} \;  \left| \mathbf{B}' \times \hat{\mathbf{n}}' \right|^{\left( p+1 \right)/2} \; \cos 2 \chi \\
\label{eq:stokes3}
U & = \kappa_{\rm s} \int {\rm d}V \; K_{\rm e,p} \; \delta^{\left( 3 + p\right)/2} \;  \left| \mathbf{B}' \times \hat{\mathbf{n}}' \right|^{\left( p+1 \right)/2}  \sin 2 \chi \; ,
\end{align}
where $\delta$ is the Doppler factor, $|\mathbf{B}' \times \hat{\mathbf{n}}'|$ is the strength of the magnetic field component perpendicular to the line of sight in the source frame,
and $\chi$ is the angle between the polarization vector of a volume element and the projection of the jet axis on the plane of the sky. The constant $\kappa_{\rm s}$, which does not affect the polarimetric observables, is defined by Eqs.~(12) and (13) of \cite{Bolis+2024}.

The integration volume is
${\rm d}V = {\rm d}z \; R\; {\rm d}R \;{\rm d}\phi$, where $0<\phi<2\pi$, $0<R<R_{0}\left(z\right)$ and $z_0<z<z_{0} + \Delta z$.
The value of $\Omega R_0$ is determined from the condition $\Gamma\left(R_0 \right)=10$. The value of $\Omega z_0$ is determined from the equation for the shape of the jet, which is $\Omega R_{0}(z_0) = 3^{1/4} (\Omega z_0)^{q}$. Finally, the parameter $\Delta z $ corresponds to the width of the emission region.

\section{Viewing angles}
\label{sec:angle}

\begin{figure*}
    \centering
    \includegraphics[width=0.4\linewidth]{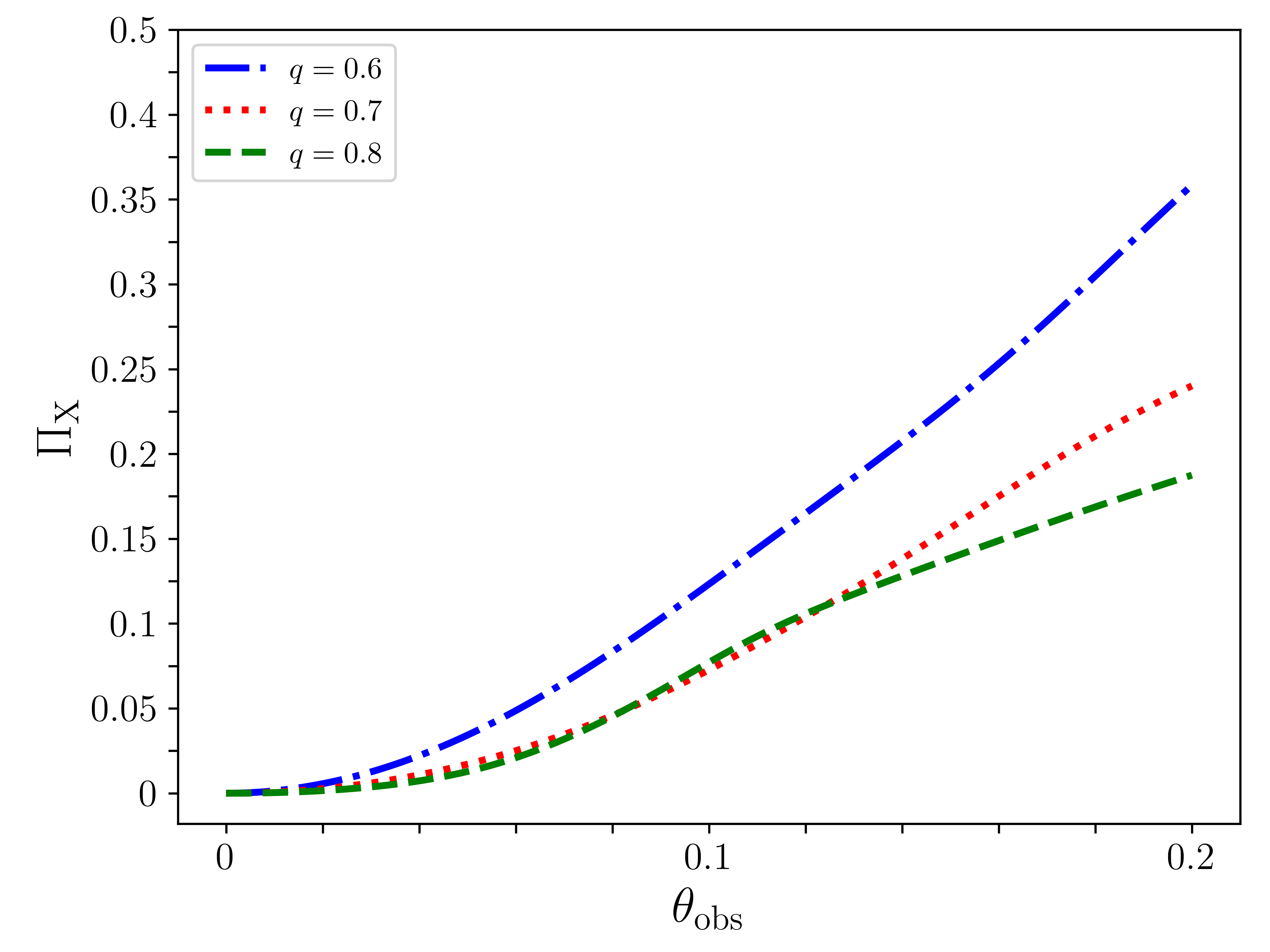}
      \includegraphics[width=0.4\linewidth]{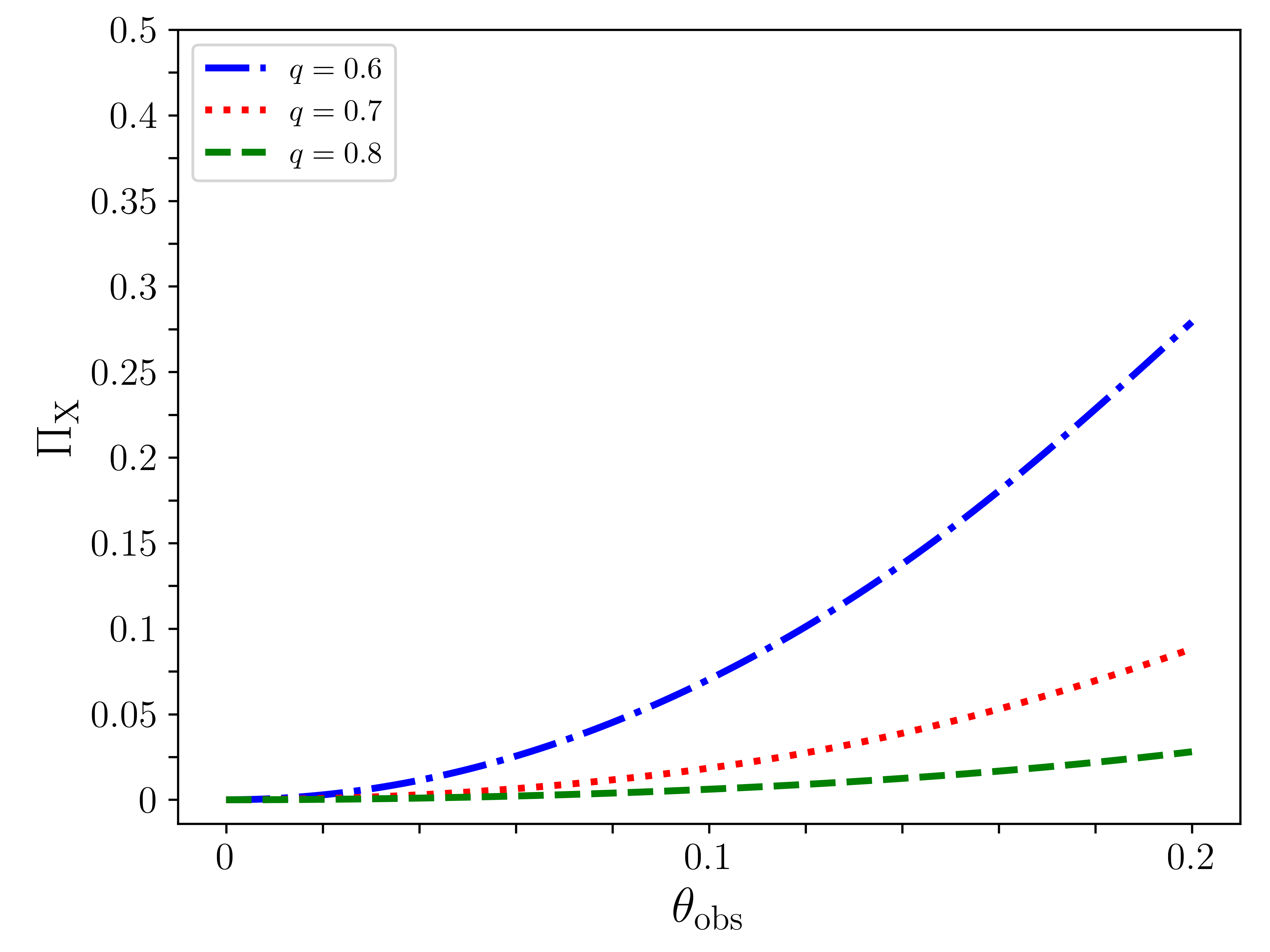}
      \includegraphics[width=0.4\linewidth]{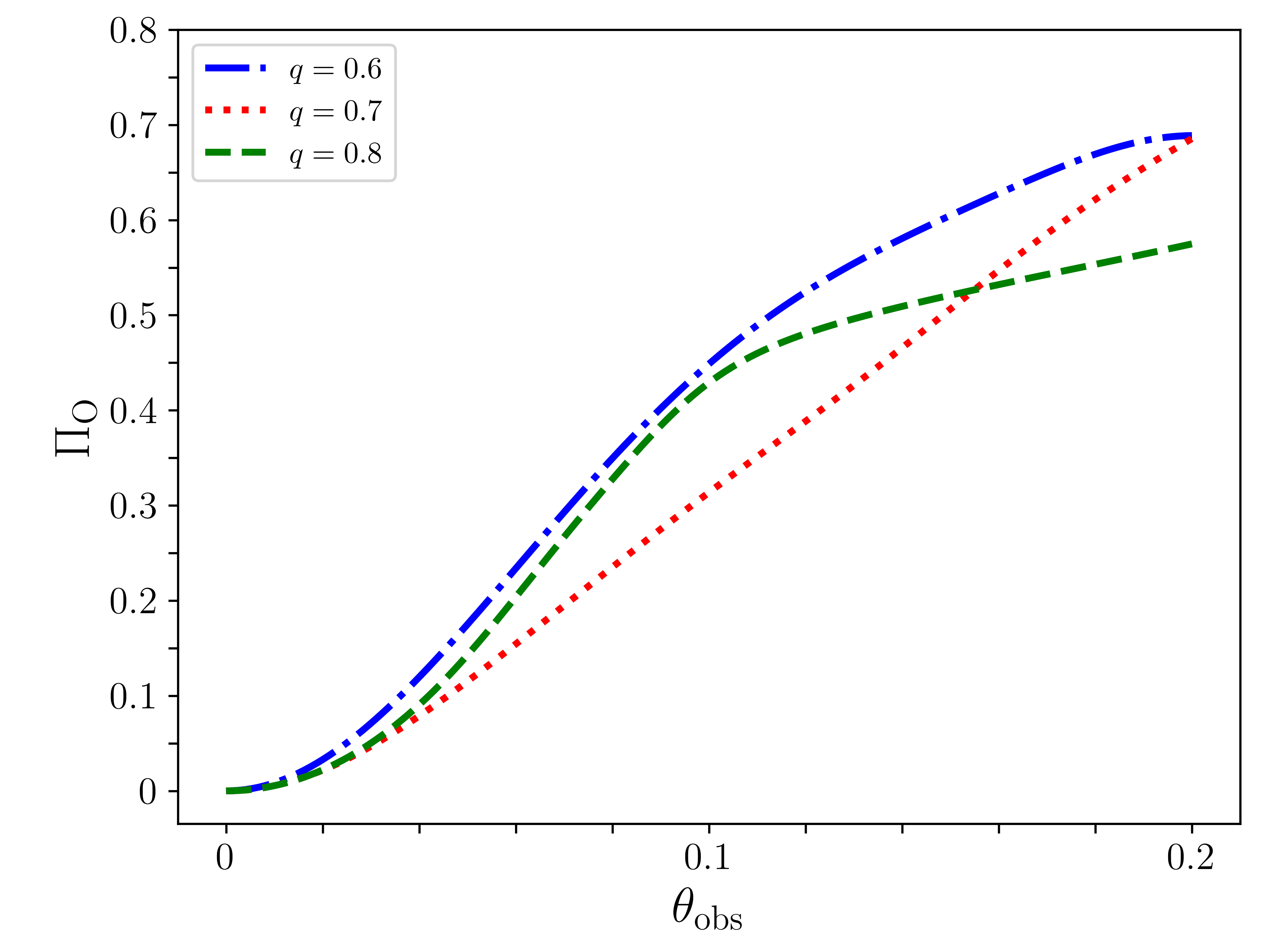}
      \includegraphics[width=0.4\linewidth]{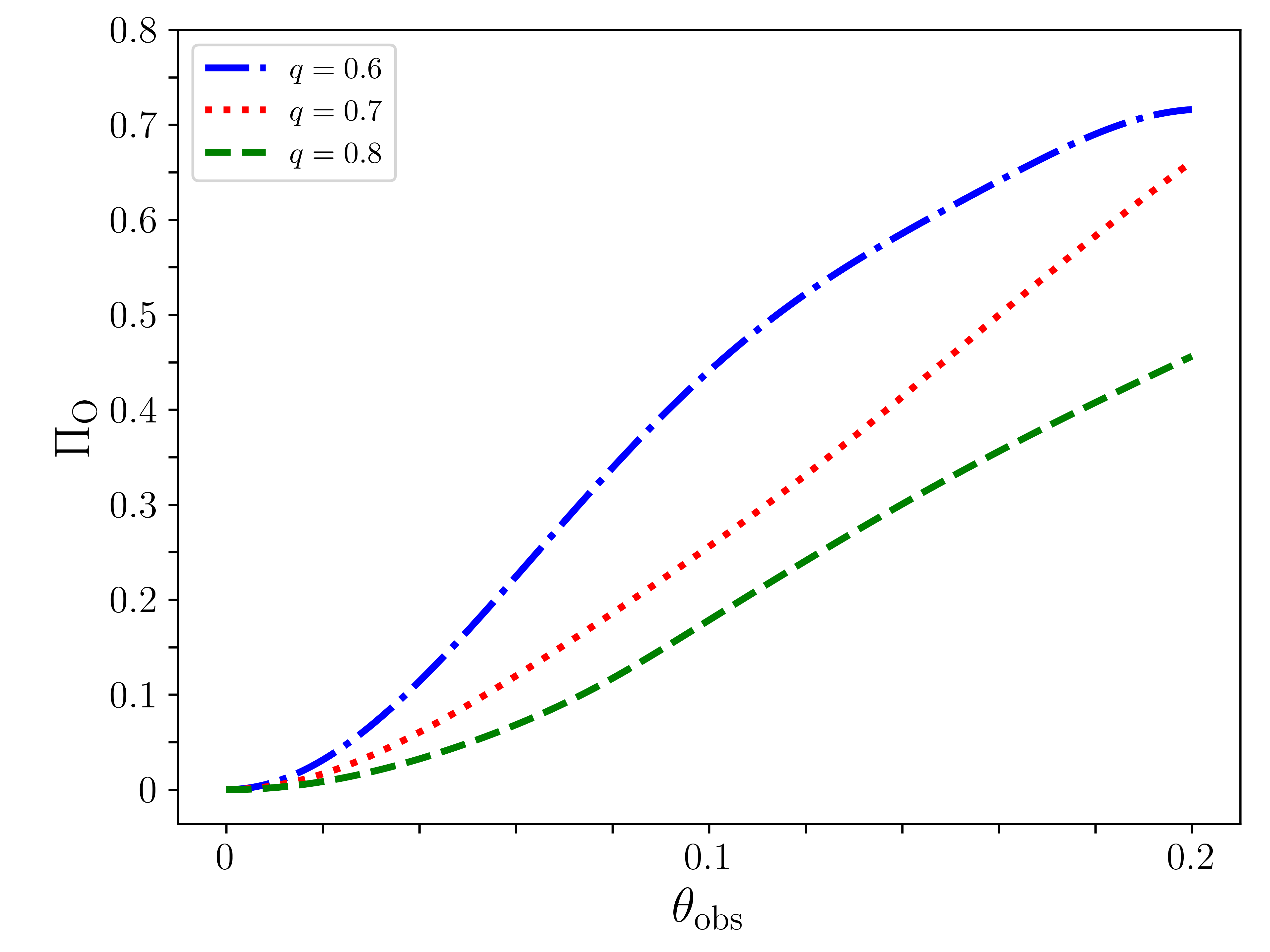}
    \caption{X-ray polarization degree, $\Pi_{\rm X}=\Pi_{p_{\rm p}=2.3}$ (top panels) and optical polarization degree, $\Pi_{\rm O}=\Pi_{p_{\rm e}=4.6}$ (bottom panels) as a function of the viewing angle, $\theta_{\rm obs}$, which is expressed in radians. In the left panels the proper number density is assumed to be constant, whereas in the right panels the magnetization is assumed to be constant. The width of the proton emission region is assumed to be $\Delta z = 3 z_0$.}
    \label{fig:angles}
\end{figure*}

In Fig.~\ref{fig:angles} we show the dependence of the degree of polarization on the viewing angle.

\end{document}